\documentclass{emulateapj}

\begin{document}
\def\mpch {$h^{-1}$ Mpc} 
\def\kms {km s$^{-1}$} 
\def\lcdm {$\Lambda$CDM } 
\def\xir {$\xi(r)$}
\def\wprp {$w_p(r_p)$}
\def\xisp {$\xi(r_p,\pi)$}
\def\xis {$\xi(s)$}
\def\rr {$r_0$}

\title{The DEEP2 Galaxy Redshift Survey: Clustering of Groups and
  Group Galaxies at $z\sim1$}
\author{Alison L. Coil\altaffilmark{1}, 
Brian F. Gerke\altaffilmark{2}, 
Jeffrey A. Newman\altaffilmark{3},
Chung-Pei Ma\altaffilmark{1},
Renbin Yan\altaffilmark{1}, 
Michael C. Cooper\altaffilmark{1}, 
Marc Davis\altaffilmark{1,2}, 
S.~M. Faber\altaffilmark{4},
Puragra Guhathakurta\altaffilmark{4}, 
David C. Koo\altaffilmark{4},
}
\altaffiltext{1}{Department of Astronomy, University of California,
Berkeley, CA 94720 -- 3411} 
\altaffiltext{2}{Department of Physics, University of California,
Berkeley, CA 94720 -- 3411} 
\altaffiltext{3}{Hubble Fellow; Institute for 
Nuclear and Particle Astrophysics, Lawrence Berkeley National Laboratory, 
Berkeley, CA 94720} 
\altaffiltext{4}{University of California Observatories/Lick
Observatory, Department of Astronomy and Astrophysics, University of
California, Santa Cruz, CA 95064}

\begin{abstract}

We study the clustering properties of groups and of galaxies in groups 
in the DEEP2 Galaxy Redshift Survey dataset at $z\sim1$ in
three separate fields covering a total of 2 degrees$^2$.  Four
measures of two-point clustering in the DEEP2 data are presented: 1)
the group correlation function for 460 groups with
estimated velocity dispersions of $\sigma\ge200$ \kms, 2) the
galaxy correlation for the full DEEP2 galaxy sample, using
a flux-limited sample of 9800 objects between $0.7\leq z\leq1.0$, 3)
the galaxy correlation for galaxies in groups or in the
field, and 4) the group-galaxy cross-correlation function.  Our
results are compared with mock group and galaxy catalogs produced from
\lcdm simulations.  Using the observed number density and clustering
amplitude of the DEEP2 groups, the estimated minimum group dark matter
halo mass is $M_{min}\sim 6 \times 10^{12} h^{-1} M_{\Sun}$ for a flat
\lcdm cosmology with $\sigma_8=0.9$.  Groups are more clustered than
galaxies in the DEEP2 data, with a relative bias of $b=1.17 \pm0.04$ 
on scales $r_p=0.5-15$ \mpch. Galaxies in groups are also more
clustered
than the full galaxy sample, with a scale-dependent relative bias 
which falls from $b\sim2.5 \pm0.3$ at $r_p=0.1$ \mpch \ to $b\sim1
\pm0.5$ at $r_p=10$ \mpch.  The correlation function of
galaxies in groups has a steeper slope ($\gamma\sim2.12 \pm0.06$) than 
for the full galaxy sample ($\gamma\sim1.74 \pm0.03$), and 
both samples can be fit by a power-law 
on scales $r_p=0.05-20$ \mpch.  We empirically measure
the contribution to the projected correlation function, \wprp,
for galaxies in groups from a `one-halo' term and a `two-halo' term
by counting pairs of galaxies in the same or in different groups.  The
projected cross-correlation between group centers and the full galaxy
sample, which is sensitive to the radial distribution of galaxies in and around
groups, shows that red galaxies are more centrally concentrated in
groups than blue galaxies at $z\sim1$.  DEEP2 galaxies in groups 
 appear to have a shallower radial distribution than that of 
 mock galaxy catalogs made from N-body simulations,
which assume a central galaxy surrounded by satellite galaxies with 
an NFW profile.  Using
simulations with different halo model parameters, we show that the
clustering of galaxies in groups can be used to place tighter 
constraints on the halo model than can be gained from using just the
usual galaxy correlation function alone.

\end{abstract}


\section{Introduction}

Groups of galaxies populate an intermediate range in density-contrast 
between galaxies and clusters and occupy a 
regime that is critical to understanding hierarchical galaxy formation
in \lcdm models. 
Merger events between galaxies likely occur within
groups rather than clusters due to their lower velocity dispersions
\citep[e.g.,][]{Ostriker80, Barnes85}. 
Galaxy groups should also be more easily related to dark matter halos
than galaxies themselves, which can have a complicated halo-occupation
function that depends significantly on galaxy properties.  
In order to understand and test galaxy formation and evolution models
it is useful to relate galaxies to observable groups as a proxy for
their parent dark matter halos. 
The current halo model paradigm 
\citep[e.g.,][]{Seljak00,Ma00,Peacock00,Cooray02,Kravtsov04} 
provides a statistical analytic measure for relating galaxies to their 
dark matter halos.  A key statistic in the halo model is the halo
occupation distribution (HOD) which measures the probability of a halo
of a given mass hosting $N$ galaxies.  
The halo model also naturally explains the small deviations seen in the 
clustering of galaxies at $z\sim0$ from a power-law model, where 
there is a transition from galaxies within a single halo and between 
different halos.  The clustering of groups and group galaxies
depends not only on halo model parameters but also on the nature of bias and 
the details of hierarchical structure formation, as well as 
cosmological parameters such as $\Omega_{m}$ and $\sigma_8$.  

Observationally, clusters of galaxies have been shown to be very
strongly clustered \citep{Bahcall88}, with the clustering strength
depending on the richness of the cluster.  \cite{Kaiser84} show that
the large clustering scale-length of massive and rare Abell clusters
can be explained by a simple model in which these clusters formed in
regions where the primordial density enhancement was unusually high.
Objects forming in the densest peaks would naturally be biased tracers
of the underlying dark matter field such that massive clusters would
have a higher correlation amplitude than that of galaxies.  Since
their masses are intermediate, galaxy groups are therefore expected to
have clustering properties between those of galaxies and clusters.

The first papers analyzing the clustering of groups in local redshift
surveys at $z\sim0$ present conflicting results and were hampered by small
samples and cosmic variance \citep[e.g.,][]{Jing88, Maia90,
Ramella90}.  \cite{Trasarti97} investigated the effect of changing the
linking-length parameters in the Friends-of-Friends (FoF) algorithm
and its effect on the clustering signal and found that these early
papers used too large a linking length, which led to a diminished
clustering strength due to the presence of interlopers.
\cite{Trasarti97} found in their data from the Perseus-Pisces redshift
survey, using two fields and $\sim50$ and 200 groups in each field,
that groups are approximately twice as clustered as galaxies, but with
significant error bars.  All of these papers showed that determining the
clustering properties of groups is a tricky endeavor, which can depend
quite sensitively on the volume and magnitude depth of
the survey, the handling of the selection function and varying
completeness, and the method used to identify groups.  In addition,
these analyses all suffered from significant uncertainties, from both
Poisson statistics, due to the small number of groups in the surveys,
and cosmic variance, due to the small volumes surveyed, which was not
quantified in any of these papers.

Significant advances have recently been made with much larger datasets.
\cite{Girardi00}, with the combined CfA2 and SSRS2 surveys, have a
sample of 885 groups in a volume of $\sim3 \times 10^5 h^{-3} 
{\rm Mpc}^3$, much larger than earlier surveys.  With this large
sample size, they are able to construct volume-limited subsamples and
investigate the dependence of clustering on group properties, finding
that groups with more members and/or larger internal velocity
dispersion are more strongly clustered.   
\cite{Merchan00} use a sample of 517 groups from the Updated Zwicky
Catalog and 104 groups from the SSRS2 to show that groups are at least
twice as clustered as galaxies, and that more massive groups have
larger clustering strength.

Most recently, the 2dF Galaxy Redshift Survey has provided a
 vast dataset with which to study large-scale
structure at $z\lesssim0.2$. Using data from the 100k release, \cite{Zandivarez03} 
measure the clustering of groups as a function of virial mass and 
find that more massive groups are more clustered and that their
measurements match the clustering of dark matter halos in a \lcdm 
N-body simulation.  
Using the completed 2dF survey, \cite{Padilla04} analyze group
clustering as a function of luminosity and show that while the least
luminous groups actually cluster less than the galaxies in the survey,
there is a strong relation between group luminosity and correlation
length, and the most luminous groups (with $L\sim4\times10^{11} h^{-2} 
L_\sun$) are $\sim10$ times more clustered
as the least luminous groups (with $L\sim2\times10^{10} h^{-2} L_\sun$).  
The relation between clustering scale
length and mean group separation that they find continues the trend
seen on larger scales for clusters.  They find very good agreement between
their data and mock catalogs constructed from
\lcdm simulations and semi-analytic galaxy evolution recipes.
These same conclusions are reached by \cite{Yang05a}, who
also use the 2dF data to measure the clustering of groups as a
function of luminosity.  It appears that at $z\sim0$ there is now convergence
among group clustering analyses.  

The extensive 2dF group catalogs have now also allowed studies of groups
 beyond simple measures of the correlation function of groups.  
Several authors  
measure the radial profile of galaxies in groups using 2dF data 
\citep{Collister05, Diaz05, Yang05b}, and analyze the cross-correlation 
between group centers and galaxies \citep{Yang05b}.  These papers find
that galaxies in groups are less concentrated than dark matter
particles in simulations, and also find that, locally, the centers of groups
are preferentially populated by red galaxies compared to blue galaxies.   
The HOD has now also been measured directly
 at $z\sim0$ using counts of galaxies in groups of different masses 
\citep{Collister05,Yang05HOD}, constraining halo model parameters locally.  
Clustering measures of the correlation function of all galaxies (not
 just those in groups) at intermediate redshifts indicate that the HOD does not 
 change significantly between $z\sim1$ and $z\sim0$ \citep{Yan03b, Phleps05}.

These measurements have only been performed with local samples; group
catalogs have not been available at intermediate- or high-redshift.
In this paper we present the first analysis of group clustering at
$z\sim1$, using group catalogs from the DEEP2 Galaxy Redshift Survey
\citep{Gerke05}.  The high resolution of the DEEP2 data allows us to
identify groups in three dimensions regardless of their galaxy
properties, using only their overdensity in space.  
We focus here on the clustering of groups and galaxies within groups at
$z\sim1$, as these measures, when combined with similar measures at
$z\sim0$, will provide constraints on galaxy evolution and structure 
formation models.  We also show how these measures can constrain
the HOD at $z\sim1$.

Four measures of two-point clustering in the DEEP2 dataset between
$0.7\leq z\leq1.0$ are analyzed: 
1) the group correlation function, 
2) the galaxy correlation function for DEEP2 galaxies,  
3) the galaxy correlation function for galaxies in groups, and 
4) the group-galaxy cross-correlation function.
These clustering measures can be used to constrain the halo model parameters 
 by comparing the data to mock catalogs with different HODs.
The first clustering measure, when combined with the observed number
density of groups in our sample, is used
to estimate the typical dark matter masses of the halos the groups
studied reside in.  The second measure provides constraints on the HOD, 
though not in an entirely unique
way; further constraints on the HOD are provided by the last two 
measures.
For the clustering of galaxies in groups, we empirically distinguish
between `one-halo' and `two-halo' terms using pairs of galaxies 
within the same group and pairs in different groups, respectively.  
The clustering of galaxies
in groups and the group-galaxy cross-correlation
function can also constrain the radial distribution of galaxies within
groups when compared with mock catalogs.

Lastly, groups may be used to constrain cosmological 
parameters like the dark energy equation of state, $w$, if their 
bivariate distribution in redshift $z$ and velocity dispersion $\sigma$ 
can be accurately measured \citep{Newman02}.  This test, however, requires 
that the relation between group velocity dispersion and dark matter halo 
mass be known and accurately calibrated. It has recently been 
suggested \citep{Majumdar04, Lima04} that the clustering properties of galaxy 
clusters may be used for "self-calibration", since the clustering 
properties of halos can be predicted as a function of mass and compared 
to the measured clustering. The group correlation
function results presented here should be useful as such a
self-calibration procedure for future DEEP2 studies.  

An outline of the paper is as follows: \S 2 briefly describes 
the DEEP2 Galaxy Redshift Survey and the sample of galaxy
groups used here, as well as the mock galaxy catalogs constructed for
the survey.  \S 3 discusses the methods used to calculate the
two-point correlation functions.  We present clustering results for
groups in \S 4, where we compare with simulations and estimate
the minimum dark matter halo mass for our groups.  In \S 5 we
analyze the clustering of the full galaxy sample and galaxies in groups in the 
DEEP2 data and in mock catalogs and show the contribution to the correlation
function for galaxies in groups from the `one-halo' and `two-halo' terms.  
\S 6 presents the cross-correlation between the full galaxy sample and
group centers, which depends upon the radial profile of galaxies in groups.
The relative biases between galaxies in groups and all galaxies 
and between groups and galaxies are presented in \S 7.
Mock catalogs with different halo models are used to 
illustrate how these various clustering measures can be used to 
constrain parameters of the halo model in \S 8. We conclude in \S 9.

\section{Data Sample and Mock Catalogs}

In this paper we use data from the DEEP2 Galaxy Redshift Survey, which is 
an ongoing project using the
DEIMOS spectrograph \citep{Faber02} on the 10-m Keck II telescope to
survey optical galaxies at $z\simeq1$ in a comoving volume of
approximately 5$\times$10$^6$ $h^{-3}$ Mpc$^3$.  Using $\sim1$~hr
exposure times, the survey will measure redshifts for $\sim40,000$
galaxies in the redshift range $0.7\sim z\sim1.5$ to a limiting magnitude
of $R_{\rm AB}=24.1$ \citep{Coil03xisp, Faber05}.  Spectroscopic
targets are pre-selected using a color cut in  $B-R$ - $R-I$ space to ensure
that most galaxies lie beyond $z\sim0.75$.  This color-cut results in a
sample with $\sim$90\% of the targeted objects at $z>0.75$, missing only
$\sim$3\% of the $z>0.75$ galaxies which meet our magnitude limit
\citep{Davis02}. Due to the high dispersion ($R\sim5,000$) of our spectra, 
our redshift errors, determined from repeated observations, are $\sim30$km
s$^{-1}$.  Restframe $(U-B)_0$ colors have been derived as described in 
\citep{Willmer05}.  Details of
the observations, catalog construction and data reduction can be found
in \citet{Davis02, Coil04, Davis04, Faber05}.

The completed survey will cover 3 square degrees of the sky over four
widely separated fields to limit the impact of cosmic variance.  Each
field is comprised of two to four contiguous photometric 'pointings'
of size $0.5$ by $0.67$ degrees.  Here we use data from six of our
most complete pointings to date, in three separate fields. 
We use data from pointings 1 and 2 in the DEEP2 fields 2, 3 and 4.
Each DEEP2 pointing corresponds to a volume of comoving dimensions
$\sim20 \times 27 \times 550$ \mpch \ in a \lcdm \ model for
$0.7\leq z \leq1.0$.  The total volume of the sample used in this paper is
thus $\sim1.8 \times 10^6 \ h^{-3} {\rm Mpc}^3$.  
To convert measured redshifts to
comoving distances along the line of sight we assume a flat \lcdm cosmology
with $\Omega_{\rm m}=0.3$ and $\Omega_{\Lambda}=0.7$.  We define $h \equiv
{\rm {\it H}_0/(100 \ km \ s^{-1} \ Mpc^{-1}})$ and quote correlation  
lengths, \rr, in comoving \mpch.
Throughout the paper, we quote empirical errors calculated from the
variance across the six pointings.  
In so doing we have treated the six DEEP2 pointings as being entirely 
independent, and they are not: there are two adjacent pointings in each of
three independent fields.  We estimate from Monte Carlo simulations using 
the methods outlined in \cite{Newman02} that, for the distribution of
pointings used here, the measured standard error should be increased by
$17 \pm5$\%, for errors which are dominated by cosmic variance; this is
a conservative assumption, as (especially for the group-group correlations) 
other contributions such as shot noise are significant.

A description of the methods employed to detect groups is presented in 
\cite{Gerke05}, along with details of the group catalog.   
A Voronoi-Delaunay Method (VDM) group-finder \citep{Marinoni02} is used to 
identify galaxy groups.  This method searches for galaxy overdensities 
in redshift space, using an asymmetric search window to account for 
redshift-space distortions.  The  advantage of the method over 
traditional FoF group-finding methods lies in the 
fact that it has no fixed length scale, but instead uses an adaptive search 
radius based on estimated group richness.  The VDM group-finder thus 
avoids a common problem of FoF methods, in which clustered groups are 
merged together along filaments.   

We use three galaxy samples in this paper; all are drawn from the same volume
as the group sample.
The full galaxy sample includes a total of 9787 galaxies 
in the redshift range $0.7\leq z \leq1.0$ in the same six pointings 
used for the group analysis.  We also split the full galaxy sample into 
subsamples of field and group galaxies, with 5947 and 3840 objects in each.
Throughout this paper we use the terms ``all galaxies'' and 
``the full galaxy sample'' interchangeably; they include both the field and 
group galaxy samples.  

Our main group sample consists of all DEEP2 groups with an 
estimated velocity
dispersion $\sigma\ge200$ \kms, with a total of 460 groups and
$\sim50-100$ groups per pointing.  We do not make a distinction
between groups and clusters - clusters are simply the larger groups;
they are included in all analyses, but have minimal impact on this work
due to their rarity.
The $\sigma\ge200$ \kms \ group sample has similar completeness ($74
\pm5$\%) and purity ($57 \pm3$\%) as samples with higher $\sigma$
cutoffs, as estimated using mock catalogs described below.  
Here completeness is defined as the fraction of real groups
(defined as a group of galaxies belonging to the same parent halo, 
where halos are identified using FoF in real space on the dark matter 
particles in the simulations)
that are successfully identified in the recovered group catalog created
by the group-finder, and
purity is the fraction of recovered groups that correspond to real
groups (see \cite{Gerke05} for more details).  As shown in Fig. 8 of
\cite{Gerke05}, these statistics are nearly independent of group
velocity dispersion.  There are more recovered groups than real groups
in these mock catalogs by a factor of 1.4. 
The $\sigma\ge200$ \kms \ 
group sample has a relatively high ``galaxy
success rate'' of $70 \pm1$\%, defined as the fraction of galaxies in
real groups that are identified as group galaxies in our recovered
group catalog, and an interloper fraction of $43 \pm1$\%.  We do not
use the $\sigma\ge350$ \kms \ cut that was used in \cite{Gerke05}, as
that cut was used to define a sample that recovered different physical
properties than the ones relevant here; 
instead of the distribution of groups in redshift and velocity 
dispersion, the relevant parameters here are the positions of groups and 
the interloper fraction (the fraction of identified group galaxies that 
are not actually in groups). 

\begin{figure}
\centerline{\scalebox{0.6}{\rotatebox{90}{\includegraphics{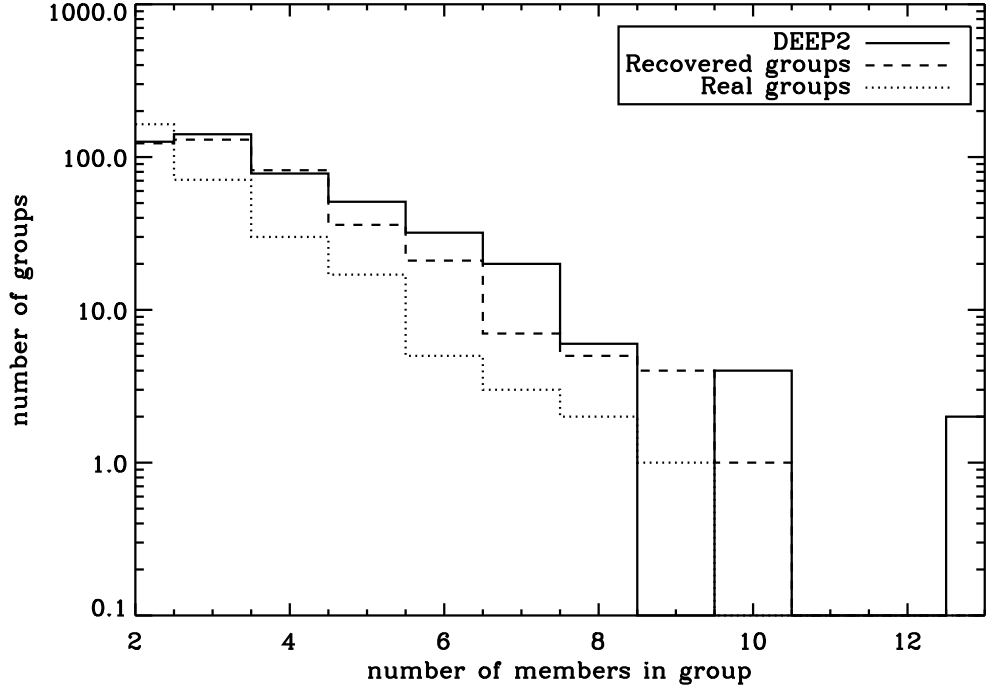}}}}
\caption{Observed richness of groups in six pointings in the DEEP2 data 
and in mock catalogs, for both real and recovered groups.  The distributions
roughly decrease as a power-law, with the largest groups having $\sim$10 
members.  There are significantly more recovered groups than real groups 
in the mock catalogs.
\label{mockrich}}
\end{figure}

The observed richness distribution of our groups in the DEEP2 data is
roughly a power-law, with most groups having two observed galaxies;
the largest groups have $\sim10$ galaxies, though there are only a few
groups this large.  Fig. \ref{mockrich} shows the observed richness
distribution 
for groups in the DEEP2 data as well as for real and recovered groups in
our mock catalogs, for groups with $0.7\leq z \leq1.0$ and $\sigma\ge200$ \kms.

\begin{figure}
\centerline{\scalebox{0.6}{\rotatebox{90}{\includegraphics{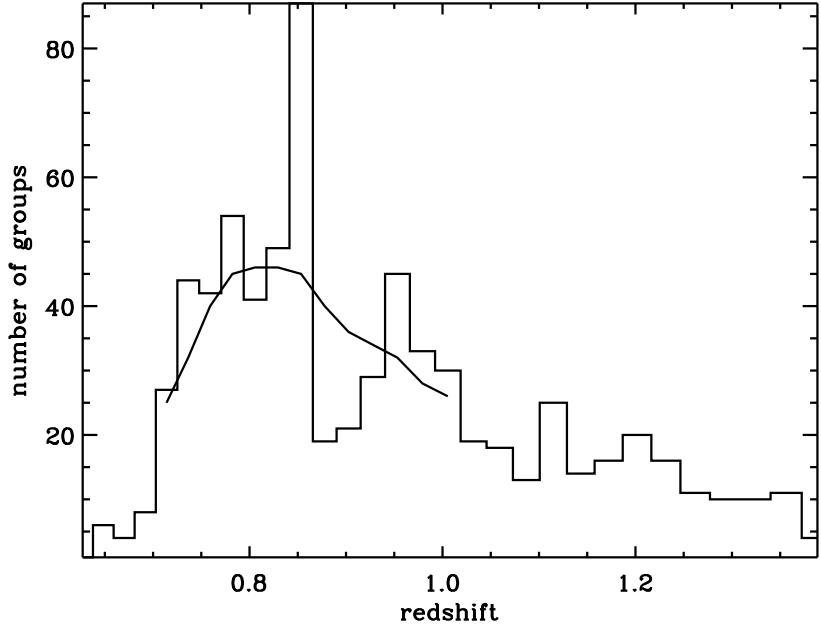}}}}
\caption{Redshift histogram of groups with $\sigma\ge200$ \kms \ 
in the DEEP2 data, using our six most complete pointings to date.  
The solid line is a smoothed version of the histogram which is used to 
estimate our selection function as a function of redshift for groups 
in our data; for the clustering analyses presented here we only use 
groups and galaxies in the redshift range $0.7\leq z \leq1.0$.
\label{sf}}
\end{figure}

\begin{figure}
\centerline{\scalebox{0.6}{\rotatebox{90}{\includegraphics{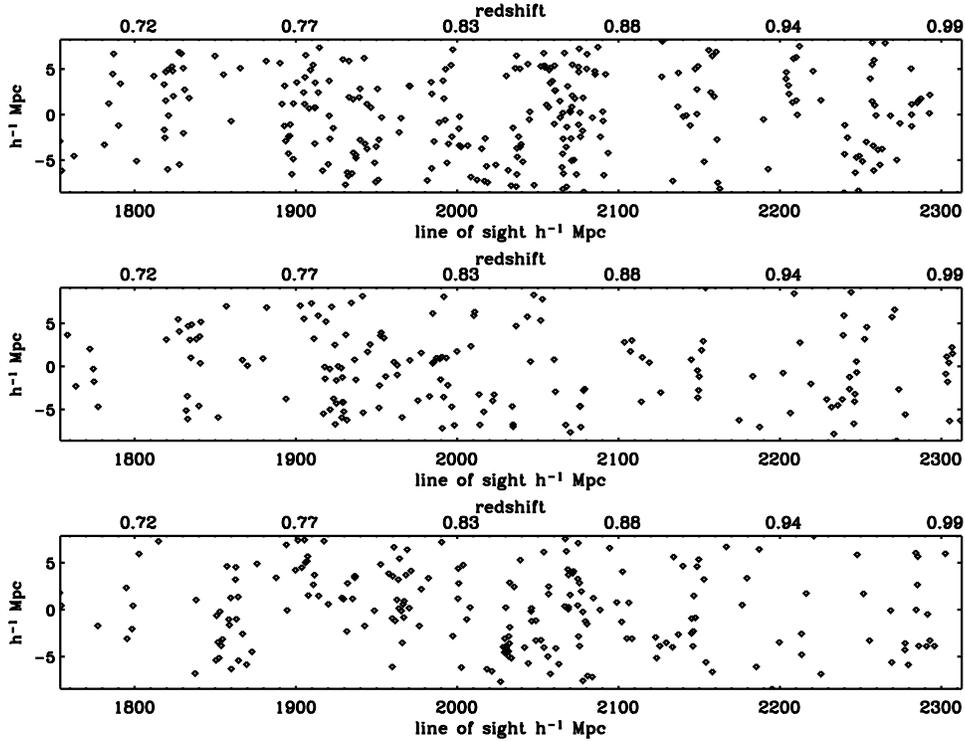}}}}
\caption{Spatial distribution of groups with  $\sigma\ge200$ \kms in three 
of the six DEEP2 
pointings used, as a function of redshift (or comoving line of sight 
distance) for $0.7\leq z \leq1.0$ and transverse distance.  We have projected 
through the short dimension of each field, corresponding to 
$\sim20$ \mpch.  The aspect ratio is not true in this figure, such that 
large-scale features appear compressed along the line of sight.  The groups
generally trace the largest structures and are often seen along filaments.
\label{conegroup}}
\end{figure}

A histogram of the redshift distribution of our group sample is shown
in Fig. \ref{sf}.  We restrict the analyses here to the redshift range
$0.7\leq z\leq1.0$ to minimize systematic effects.  While groups are
found in the survey to higher redshifts, those groups are likely to be more
massive and hence not as representative of groups in our sample as a
whole.  Additionally, the $R$-band target selection of the survey
corresponds to a bluer restframe color-selection at higher redshift;
this results in fewer red galaxies being
targeted at higher redshift compared to blue galaxies \citep{Willmer05}, 
which could systematically affect the group richness and velocity
dispersion estimates at $z\gtrsim1.0$.
The spatial distribution of DEEP2 groups in three of the six pointings
used is shown in Fig. \ref{conegroup}.  Shown are pointings 2 in the DEEP2 
fields 2, 3 and 4.  

The mock galaxy catalogs used throughout the paper are described in 
\cite{Yan03}; relevant
details are repeated here.  The mock catalogs are constructed from
N-body simulations of $512^3$ dark matter particles with a particle
mass of $m=1 \times 10^{10} h^{-1}$ $M_{\Sun}$ in a box with dimensions
256 \mpch \ on a side, for a \lcdm \ cosmology with $\sigma_8=0.9$.
Dark matter halos were identified using FoF in real space and 
galaxies were placed within halos using a halo model prescription.
To populate dark matter halos with galaxies, two functions 
need to be specified. The first is the halo occupation distribution function 
(HOD), which is the probability that a halo of mass M hosts  
N galaxies, $P(N|M)$. The second is the spatial distribution of galaxies 
within halos.  
The first moment of the HOD function, the average number of galaxies as a 
function of halo mass M, is shown in Fig. 1 in \cite{Yan03} and in
Fig. 12 of this paper (labeled as ``B256'')
for the HOD used in the mock galaxy catalogs here. This HOD is 
constrained at $z\sim 0$ with the 2dF luminosity function
and luminosity dependent two-point correlation function of galaxies 
and at $z\sim 1$ with the DEEP2 $\xi(r)$ and the COMBO-17 luminosity functions.
In \S 8 of this paper we compare clustering results for 
mock catalogs with two different halo models; for 
the bulk of the paper, however, we use lightcones with the HOD as
published in \cite{Yan03}.  
Once the number of galaxies in each halo and their corresponding
luminosities are known, the most luminous galaxy is assigned to the
center of mass of the halo, and positions and velocities for the 
other galaxies are drawn randomly from those of the dark matter particles.  
No other radial or velocity bias is 
included in assigning galaxies to particles, so that galaxies trace 
the mass and velocity distributions
of the dark matter particles in the halo. 
The spatial distribution of galaxies follow the dark matter
density profile, which on average is an NFW profile. 
Although no velocity bias is included, the velocity dispersions of the
galaxies are systematically smaller than those of the mass (see Fig.8
in Yan, White \& Coil 2004), due to the fact that the most luminous
galaxy is always assigned to the particle at the center of mass.
Here we use a set of twelve independent catalogs which each have the
same spatial extent as a single DEEP2 pointing ($\sim0.5$ by $0.67$
degrees).

The center of each DEEP2 group (needed for the clustering measures
performed here) is measured as the median of the positions (in
comoving x, y, and z) of the galaxies identified in that group.
Errors on the positions of the recovered groups are estimated using
the differences in the mock catalogs between the centers of real
groups and recovered groups.  As stated above, there is a factor of
1.4 more recovered groups in the mock catalogs than real groups for 
$\sigma\ge200$ \kms.  In
the mock catalogs 74\% of recovered groups have a real group within
$r_p=1$ \mpch, while 48\% have a real group within $r_p=0.2$ \mpch \
and 41\% have a real group within $r_p=0.1$ \mpch.  Here $r_p$ is the
projected distance on the plane of the sky, which is the relevant
distance for the projected clustering used in \S 5 and later in the
paper; in \S 4 we use the redshift-space correlation function,
but only measure it on scales $s>2$ \mpch.  We therefore believe that
on scales larger than 1 \mpch \ our results are robust to errors in
the group positions, while on smaller scales, where we will measure
the cross-correlation between group centers and galaxies, there is
likely to be some degradation in the signal due to position errors;
this is discussed more in \S 9. 
However, the mock catalogs have been treated in an identical manner as
the data so that comparisons between the data and mock catalogs are
unaffected.

\section{Methods}

The two-point correlation function \xir \ is defined as a measure of the excess
probability above Poisson of finding an object in a volume element $dV$
at a separation $r$ from another randomly chosen object,
\begin{equation}
dP = n [1+\xi(r)] dV,
\end{equation}
where $n$ is the mean number density of the object in question 
\citep{Peebles80}.  

Measuring \xir \ requires constructing a random catalog with the same
selection criteria and observational effects as the data, to serve 
as an unclustered
distribution with which to compare.  For each data sample we
create a random catalog with the same overall sky coverage 
and redshift distribution as the data.  This is achieved by first applying the
two-dimensional window function of our data in the plane of the sky 
to the random catalog.  Our overall redshift success rate 
is $gtrsim$70\% and is not entirely uniform across the survey; 
some slitmasks are observed under better conditions than others and 
therefore yield a slightly higher completeness.  
This spatially-varying redshift success rate is taken into account in 
the spatial window function which is applied to both the random
catalog and the mock catalogs, such that regions of the sky with a
higher completeness have a correspondingly higher number of random
points or more objects in the mock galaxy catalogs.  This ensures that there
is no bias introduced when computing correlation statistics or
comparing the data to the mock catalogs.  We also
mask the regions of the random and mock catalogs 
where the photometric data have saturated stars and CCD defects. 

We then apply a selection
function, $\phi(z)$, defined as the probability of observing a group as 
a function of redshift, so that the random catalog has the same overall
redshift distribution as the data.  The selection
function for groups in the DEEP2 survey is determined by smoothing the
observed redshift distribution of groups in the catalog used here, as
shown in Fig. \ref{sf}.  The apparent overdensity at $z\sim0.85$ seen in
this figure is due to structures in several pointings and does not
significantly affect our results, which are averaged over all
pointings and a wide redshift range.  As the group catalog includes 
three separate fields, the data are combined when calculating
$\phi(z)$, which reduces the effects of cosmic variance. Using a 
smoothed redshift distribution to estimate the selection function 
can cause a systematic bias, but we estimate that this bias is less 
than the errors on \xir.  In what
follows, for both the data and the mock catalogs the redshift
range over which we compute \xir \ is limited to 
$0.7\leq z \leq1.0$, the redshift 
range over which the selection function varies by less than a factor of two.
Each of the six pointings has between $\sim50-100$ groups, and each of 
the twelve independent mock catalogs 
contains $\sim100$ groups. We use 
$\sim3,000$ random points in each pointing to calculate the correlation 
function for groups and $\sim20,000$ random points to calculate the 
correlation function for galaxies.

The two-point correlation function is measured for both 
groups and galaxies using the
\citet{Landy93} estimator,
\begin{equation}
\xi=\frac{1}{RR}\left[DD \left(\frac{n_R}{n_D}
\right)^2-2DR\left(\frac{n_R}{n_D} \right)+RR\right],
\end{equation}
where $DD, DR$, and $RR$ are pair counts in a given separation range 
in the data-data, data-random, and random-random catalogs, and $n_D$
and $n_R$ are the mean number densities in the data and
random catalogs.  This estimator has been shown to perform as well as the
Hamilton estimator \citep{Hamilton93} but is preferred as it is 
relatively insensitive to the size of the random catalog and handles 
edge corrections well \citep{Kerscher00}.   As the magnitude limit of
our survey results in a non-uniform selection function,  a 
standard $J_3$-weighting scheme is applied \citep{Davis82den}, which 
attempts to
weight each volume element equally while minimizing the variance on
large scales.  As the redshift range is limited to $0.7\leq z \leq1.0$ for
all analyses in this paper, this effect is not large ($\lesssim20$\%).

Measurements of the cross-correlation between two samples (presented in \S6) 
use a symmetrized version of Eqn. 2.  Each data sample, with pair counts 
labeled $D_1$ and $D_2$, has an associated random catalog, with pair counts 
$R_1$ and $R_2$, with the same selection 
function as the data.  After normalizing each data and random catalog by its 
number density, the cross-correlation is estimated using
\begin{equation}
\xi=\frac{1}{R_1R_2}\left[D_1D_2-D_1R_2-D_2R_1+R_1R_2\right].
\end{equation}

Redshift-space distortions due to peculiar velocities along the
line-of-sight will introduce systematic effects to the estimate of
\xir. At small separations, random motions within a virialized overdensity
cause an elongation along the line-of-sight (``fingers of God''),
while on large scales, coherent infall of galaxies into forming
structures causes an apparent contraction of structure along
the line-of-sight (the ``Kaiser
effect'').  What is actually measured then is \xis, where $s$ is the
redshift-space separation between a pair of galaxies.  As we will show
in the next section using mock group catalogs, for galaxy groups over
the scales relevant here, the systematic effects of redshift-space
distortions are of order 20\%.  
Mock catalogs are used to correct for redshift-space distortions and infer \xir
\ from \xis by multiplying the observed \xis for DEEP2 
groups by the ratio of \xir to \xis for groups in the mock catalogs.

However, redshift-space distortions are more significant when measuring 
$\xi$ for galaxies, as smaller scales are probed.  
We are able to measure $\xi$ on small scales for the galaxy sample 
both because of the larger sample size, which allows us to stably measure 
the clustering on small scales, and also because there is no exclusion 
radius, unlike with groups, which typically have a radius of $R\sim1$ \mpch. 
To uncover the real-space clustering properties of galaxies we measure
$\xi$ in two dimensions, both perpendicular to and along the line of
sight.  Following \cite{Fisher94}, ${\bf v_1}$ and ${\bf
v_2}$ are the redshift positions of a pair of galaxies, ${\bf s}$ is 
 the redshift-space separation (${\bf v_1}-{\bf v_2}$), and ${\bf l}
=\frac{1}{2}$(${\bf v_1}+{\bf v_2})$ is the mean distance to the
pair.  The separation between the two galaxies across
($r_p$) and along ($\pi$) the line of sight are defined as
\begin{equation}
\pi=\frac{{\bf s} \cdot {\bf l}}{{\bf |l|}},
\end{equation}
\begin{equation}
r_p=\sqrt{{\bf s} \cdot {\bf s} - \pi^2}.
\end{equation}
In applying the \citet{Landy93} estimator to galaxies, 
pair counts are computed over a two-dimensional grid of separations to
estimate \xisp.

To recover \xir \ \xisp \ is projected along the $r_p$ axis.
As redshift-space distortions affect only the line-of-sight component
of \xisp, integrating over the $\pi$ direction leads to a statistic
\wprp, which is independent of redshift-space distortions.  Following
\cite{Davis83},
\begin{equation}
w_p(r_p)=2 \int_{0}^{\infty} d\pi \ \xi(r_p,\pi)=2 \int_{0}^{\infty}
dy \ \xi(r_p^2+y^2)^{1/2},
\label{eqn}
\end{equation}
where $y$ is the real-space separation along the line of sight. If
\xir \ is modeled as a power-law, $\xi(r)=(r/r_0)^{-\gamma}$, then
\rr \ and $\gamma$ can be readily extracted from the projected
correlation function, \wprp, using an analytic solution to Equation
\ref{eqn}:
\begin{equation}
w_p(r_p)=r_p \left(\frac{r_0}{r_p}\right)^\gamma
\frac{\Gamma(\frac{1}{2})\Gamma(\frac{\gamma-1}{2})}{\Gamma(\frac{\gamma}{2})},
\label{powerlawwprp}
\end{equation}
where $\Gamma$ is the gamma function.  A power-law fit to \wprp \ will
then recover \rr \ and $\gamma$ for the real-space correlation
function, \xir.

\section{Group Clustering Results}

This section presents results on the clustering of groups at
$z\sim1$.  The two-point correlation function for
groups in the DEEP2 data is measured.  The clustering properties
of groups in our mock catalogs are analyzed, which allows us to quantify our
systematic errors and to correct the observed clustering in the DEEP2 data
for redshift-space distortions and the effects of the group-finder.
The number density of groups in the DEEP2 data 
is then used to infer the minimum dark matter halo mass which hosts
our groups; we then compare the clustering of halos with this mass to
the clustering of our group sample.

\subsection{Clustering of Groups in DEEP2 Data and Mock Catalogs}

\begin{figure}
\centerline{\scalebox{0.7}{\rotatebox{90}{\includegraphics{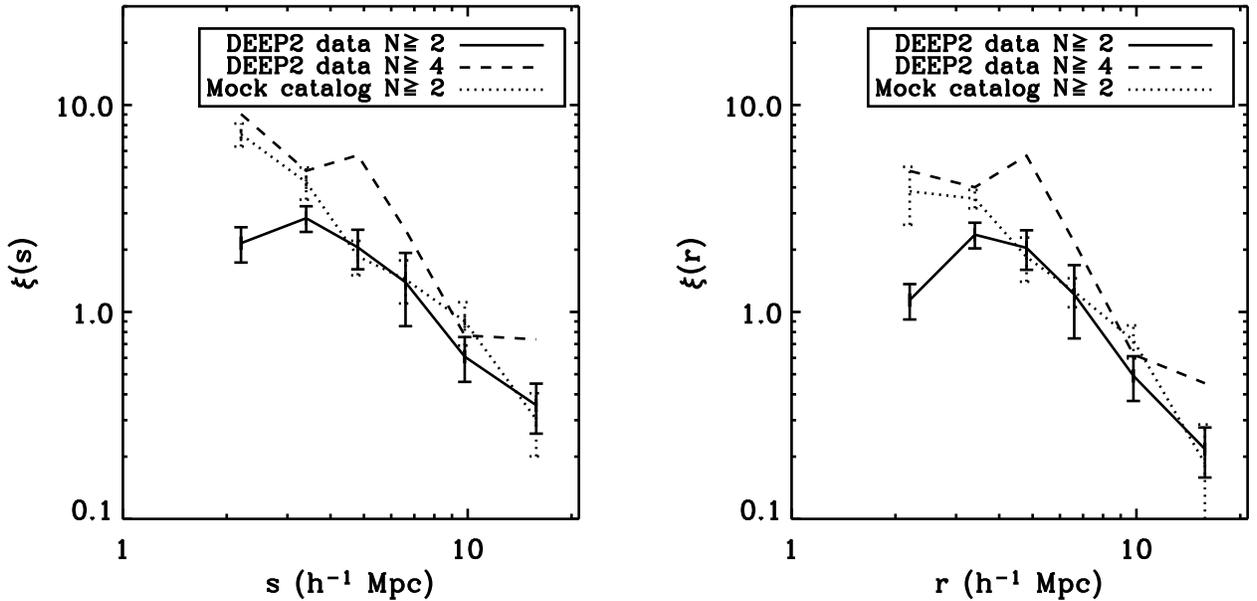}}}}
\caption{The clustering of groups with $\sigma\ge200$ \kms \ 
in DEEP2 data and mock catalogs.
The redshift-space correlation function, \xis \ (left), and real-space 
correlation function, \xir \ (right), 
 for the groups with four or more observed members ($N\ge4$, dashed line) 
are higher than for groups with two or more observed members ($N\ge2$, solid 
line).  The average \xis \ and \xir \ for groups with $N\ge2$
in the mock catalogs (dotted line) is shown as well, and agrees well
with the data.  Corrections have been made for the DEEP2 
slitmask target selection algorithm. 
\label{mocks}}
\end{figure}

The left panel of Fig. \ref{mocks} shows the observed two-point
correlation function in redshift space, \xis, for our $\sigma\ge200$ \kms 
group sample, for both the full sample (solid line) with two or more galaxies
in each group ($N\ge2$) and for a subsample with four or more galaxies in 
each group ($N\ge4$).  Also shown is \xis \ for the recovered 
group sample in the mock catalogs with $\sigma\ge200$ and $N\ge2$ 
(dotted line).  We show the standard error across the six 
pointings; this empirical error therefore 
includes both cosmic variance and Poisson error. 
The $N\ge4$ data sample shows a stronger 
clustering amplitude than the $N\ge2$ data sample.  This is likely due to the 
$N\ge4$ sample containing more massive groups, on average, than 
the $N\ge2$ sample.  The $N\ge4$ sample is also less likely to have 
interlopers, which may also increase the observed clustering.

Power-law fits to \xis \ over the range $s=3-20$
\mpch \ are given in Table 1.
Below $s\sim3$ \mpch \ \xis \ does not continue to rise as a
single power-law in our data; this lack of pairs on scales $s<3$ \mpch
\ is likely due to the finite physical extent of groups.  To take into
account the covariance between $s_0$ and $\gamma$, we perform a
$\chi^2$ minimization and marginalize over each parameter separately. 
This procedure leads to errors which are roughly a factor of two larger than
if we neglected this covariance.

The mock galaxy catalogs described in \S 2 are used to quantify 
systematic errors in our group clustering analysis.   
Effects that may be introduced by the group-finder are tested by comparing 
 \xis \ in redshift space for both real and recovered groups.
The results are shown in Table 1.
The recovered groups have a slightly higher 
clustering amplitude (5\%) than the real groups.
Redshift-space distortions are 
quantified by measuring \xis \ in redshift space and \xir \ in real space  
for real groups.  
Redshift-space distortions appear to enhance the clustering properties
of groups by $\sim20$\% on scales $s\sim2-15$ \mpch \  and
decrease the clustering amplitude on smaller scales. 
The clustering scale length increases by $\sim20$\% when measured in 
redshift space,
while there is no change to the slope over the scales used here.
The effects of our slitmask target selection algorithm on the
clustering of groups are also tested using the mock catalogs.  Our target
selection code determines which galaxies would be observed on
slitmasks; because spectra from neighboring galaxies can not overlap
on the CCD, not all galaxies can be selected to be observed.  In
particular, in overdense regions on the plane of the sky the number
of galaxies which can be observed decreases in a non-trivial way.  By
running our slitmask target selection code on the mock catalogs we
can quantify the effect on the measured \xir. 

Under the assumption that corrections for the above effects should be
proportional to the clustering strength, these results from the mock
catalogs can now be used to correct the observed \xis \ for groups in
the DEEP2 data for target selection effects, redshift space
distortions, and the group-finder, in order to estimate \xir \ in real
space for real groups.  We can either apply corrections to the
power-law fits themselves or to the data points as a function of
scale.  If we apply the corrections to $s_0$ and $\gamma$ as measured
for the DEEP2 groups, the corrections would infer that $r_0=6.8\pm0.6$
\mpch \ and $\gamma=1.4 \pm0.2$ for the real-space correlation
function of real groups in the DEEP2 data with $N\ge2$.  If we
explicitly correct the observed \xis \ for the DEEP2 groups as a
function of scale, using the ratio of \xir/\xis \ in the mock
catalogs, the resulting \xir \ has a power-law fit of $r_0=6.2 \pm0.4$
\mpch \ and $\gamma=1.5 \pm0.2$, within the 1$\sigma$ error of the
inferred values.  The corrected \xir \ is shown as the solid line in
the right panel of Fig. \ref{mocks}, for both the $N\ge2$ and $N\ge4$
samples.  The clustering of groups in the DEEP2 data is therefore
1-3$\sigma$ lower than the clustering of real groups before target
selection in the mock catalogs, where $r_0=7.4 \pm0.4$ and $\gamma=1.6
\pm0.2$, for $N\ge2$.

\subsection{Minimum Group Mass Derived from Clustering Results}

The minimum dark matter halo mass that our groups reside in can be
inferred from the observed number density and clustering strength of
our groups using either analytic formulations of the dark matter halo
mass function or by comparing directly to simulations.  Here we
estimate a minimum dark matter halo mass from the observed number
density using analytic theory and then compare the corresponding
predicted clustering strength of those halos in both theory and
simulations with the observed clustering strength of our groups.

The $\sigma\ge200$ \kms \ group sample has an observed
density of $n=4.5 \times 10^{-4} \ h^3 \ {\rm Mpc}^{-3}$, calculated from the
observed number of groups between $z=0.75-0.85$, where the group
selection function is the highest, divided by the comoving volume
occupied by the groups.  The mock catalogs are used to estimate the
effects on the observed number density due to our slitmask target
selection (which decreases the number of observed groups, as most
groups have only two or three observed galaxies) and the false
interloper rate due to the group-finder.  Correcting for these
effects, the actual comoving number density is estimated to be \
$n=6 \times 10^{-4} \ h^3 \ 
{\rm Mpc}^{-3}$.  This corresponds to a mean inter-group spacing of
$d=11.8$ \mpch.  For comparison to the sample used by \cite{Gerke05}, 
the number density of groups with $\sigma\ge350$ \kms \ is  
$n=2.4 \times 10^{-4} \ h^3 \ {\rm Mpc}^{-3}$, after applying the above
corrections, which corresponds to an intergroup spacing of $d=16$
\mpch.

For a \lcdm cosmology with $\Omega_b=0.05$, $\Lambda=0.7$,
$\Omega_m=0.3$, $h=0.7$, $\sigma_8=0.9$, a comoving number density of $n=6
\times 10^{-4} h^3 {\rm Mpc}^{-3}$  results in 
$M_{min}=5.9 \times 10^{12} h^{-1} M_{\Sun}$ at $z=0.8$ and 
$M_{min}=5.5 \times 10^{12} h^{-1} M_{\Sun}$ at $z=1$
for a \cite{Sheth99} mass function.  
A comoving density of $n=2.4 \times 10^{-4} \ h^3 \ {\rm Mpc}^{-3}$, 
estimated for the $\sigma\ge350$ \kms \ sample, corresponds to 
$M_{min}=1.2 \times 10^{13} h^{-1} M_{\Sun}$ at $z=0.8$ and 
$M_{min}=1.1 \times 10^{13} h^{-1} M_{\Sun}$ at $z=1$.
These masses are only approximate as the group number and volume are both 
just estimates.  The 
errors on the minimum dark matter halo masses inferred from the
observed number densities are likely $\sim50$\%, given cosmic variance
errors and the uncertainties in the corrections made to the number
densities from the mock catalogs.

We check for consistency between the observed and predicted clustering of 
these halos.  
\cite{Mo02} use the \cite{Sheth99} model to predict the
evolution in the clustering of dark matter halos and find that 
halos of mass $M_{min}=5.5 \times 10^{12} h^{-1} M_{\Sun}$ will have a 
clustering amplitude of $\sigma_8=1.0$ at $z=1$.  Here  
 $\sigma_8$ is defined as the standard deviation of halo count
fluctuations in a sphere of radius 8 \mpch; it 
can be preferable to quoting a scale-length, $r_0$, as it 
removes the significant covariance with $\gamma$.
$\sigma_8$ \ can be calculated from a power-law fit to \xir \
using the formula,
\begin{equation}
(\sigma_8)^2 \equiv J_2(\gamma) \left(\frac{r_0}{8 \ h^{-1} {\rm Mpc}}\right)^\gamma,
\label{sig8eqn}
\end{equation}
where
\begin{equation}
J_2(\gamma) = \frac{72}{(3-\gamma)(4-\gamma)(6-\gamma) 2^\gamma}
\end{equation}
 \citep{Peebles80}.
Using the power-law fits to \xir \ for groups in the DEEP2 data, we
find
$\sigma_8=1.0$, in agreement with the predicted value of \cite{Mo02}.

\cite{Kravtsov04} find using dark matter simulations with the same
concordance cosmology that a number density of 
$n=6 \times 10^{-4} \ h^3 \ {\rm Mpc}^{-3}$ at
$z=1$  corresponds to a minimum mass of $M_{min}=5.9 \times 10^{12}
h^{-1} M_{\Sun}$.  This value is comparable to the \cite{Sheth99}
value quoted above. They predict a clustering amplitude of $r_0=5.2$ \mpch \
and $\gamma=2.16$ for these halos, which corresponds to $\sigma_8=1.05$, 
in good agreement with our observed value of $\sigma_8$ for the DEEP2 groups.

We verify from the mock catalogs that the actual minimum 
dark matter halo masses for these group samples are similar to
those estimated above.  For the group sample
defined as having an estimated $\sigma\ge200$ \kms, the mass distribution has 
a rough minimum dark matter halo mass of
$M_{min}=2-3 \times 10^{12} h^{-1} M_{\Sun}$  and 
the $\sigma\ge350$ \kms \ sample has a rough minimum dark matter halo mass of
$M_{min}=4-5 \times 10^{12} h^{-1} M_{\Sun}$. 
The mass distributions for the group samples in the mock catalogs do not  
have a very clearly defined lower-mass cutoff, however, and these values are 
roughly half of the values quoted above.  Even though the galaxies in the 
mock catalogs are randomly drawn from the velocity distribution of 
individual dark matter particles,
the fact that only a small number of galaxies are observed in a single group
leads to significant scatter between 
the estimated $\sigma$ from the observed galaxies and the actual
$\sigma$ and therefore the mass of the underlying dark matter particles; 
this scatter likely accounts for the factor of two discrepancy between
the halos in the mock catalogs and the predictions of \cite{Sheth99}.

\section{Clustering of Galaxies in Different Environments}

In this section the clustering properties for galaxies 
in groups relative to the full galaxy population and to field galaxies are
investigated.  Unlike for the group sample, which has larger Poisson errors 
and can only be measured on scales $r\ge3$ \mpch, where redshift-space 
distortions are small, here we are interested in measuring
clustering properties of galaxies on small scales where redshift-space 
distortions are much more significant.  Instead of inferring \xir \ in 
real space from 
measurements of \xis \ in redshift space, we measure the projected 
two-point correlation function, \wprp, from which \xir \ can be more directly 
inferred.  Color information additionally allows us to divide the 
sample of group galaxies into red and blue populations and measure 
\wprp \ for each.  We test the effects of our slitmask target 
selection algorithm and group-finder on these results using mock catalogs.
Finally, we are able to empirically separate the observed \wprp \ for galaxies 
in groups into a `one-halo' and a `two-halo' term, by keeping pair 
counts where both galaxies are in the same or in different groups.

\subsection{Clustering of Full Galaxy Sample and Galaxies in Groups in DEEP2 data}

\begin{figure}
\centerline{\scalebox{0.6}{\rotatebox{90}{\includegraphics{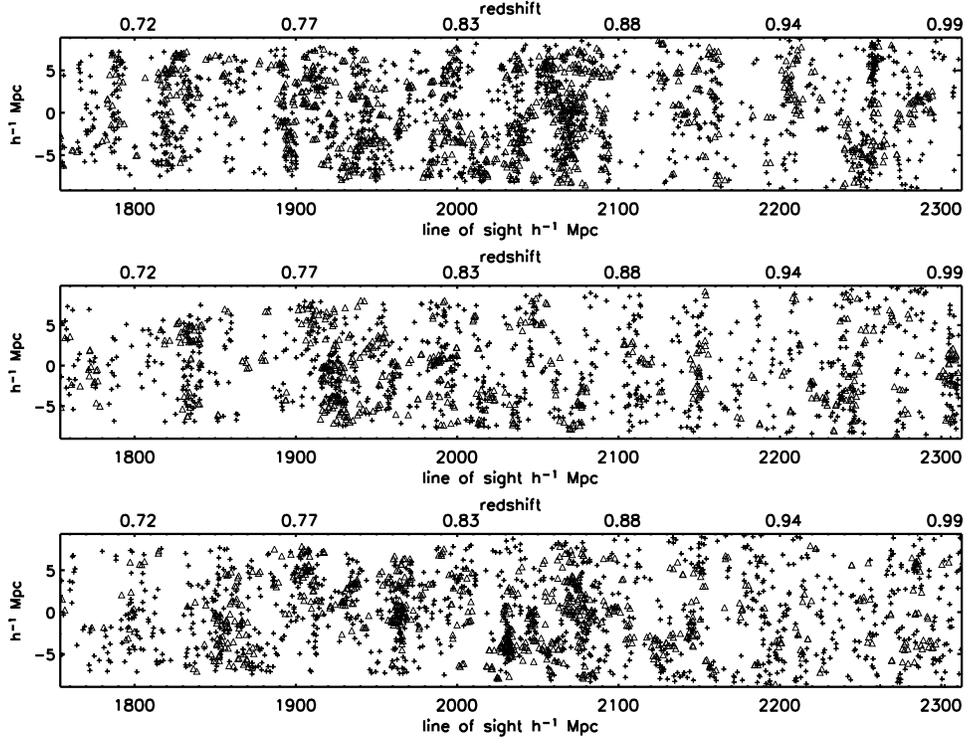}}}}
\caption{Spatial distribution of DEEP2 galaxies identified as being 
 field galaxies (thin crosses) or groups galaxies (thick stars) 
for data in the same three pointings plotted in Fig. \ref{conegroup}.  
As in Fig. \ref{conegroup}, we have projected
 through the short dimension of each field, corresponding to $\sim20$ 
\mpch.
\label{coneplot}}
\end{figure}

The full galaxy sample here refers to all DEEP2 galaxies between $0.7\leq z 
\leq1.0$ in the same six pointings for which we have group catalogs. 
The group and field galaxy samples are identified using the 
$\sigma\ge200$ \kms \ 
group catalog, and these samples combined make up the full galaxy sample. 
  Fig. \ref{coneplot} shows the spatial distribution of
galaxies in three DEEP2 pointings, with different symbols for group
(open triangles) and field (crosses) galaxies.  In the DEEP2 data 
39 +/-4\% of all galaxies are identified as belonging in recovered 
groups which have $\sigma\ge 200$ \kms \ in the redshift range 
$0.7\leq z \leq1.0$, where the error quoted is the standard error 
across the six
pointings.  This rate is artificially increased by false interlopers,
but it is also decreased by our slitmask target selection by roughly
the same amount, as estimated using mock galaxy catalogs.  In the mock
catalogs, 27\% of all galaxies are identified as belonging in 
recovered groups after target selection (24\% are in
real groups before target selection), significantly less than in the 
DEEP2 data.  This difference is most likely due to the free parameters in
the group-finding algorithm having been tuned to reproduce the
observed $n(\sigma,z)$ for $\sigma\ge350$ \kms, not the $\sigma\ge200$
\kms \ cutoff in the present sample (see Fig. 6 in \cite{Gerke05} for 
details).  The clustering results shown here do not depend on the 
absolute number of galaxies in each of the samples (all, group, and
field galaxies).

\begin{figure}
\centerline{\scalebox{0.5}{\rotatebox{90}{\includegraphics{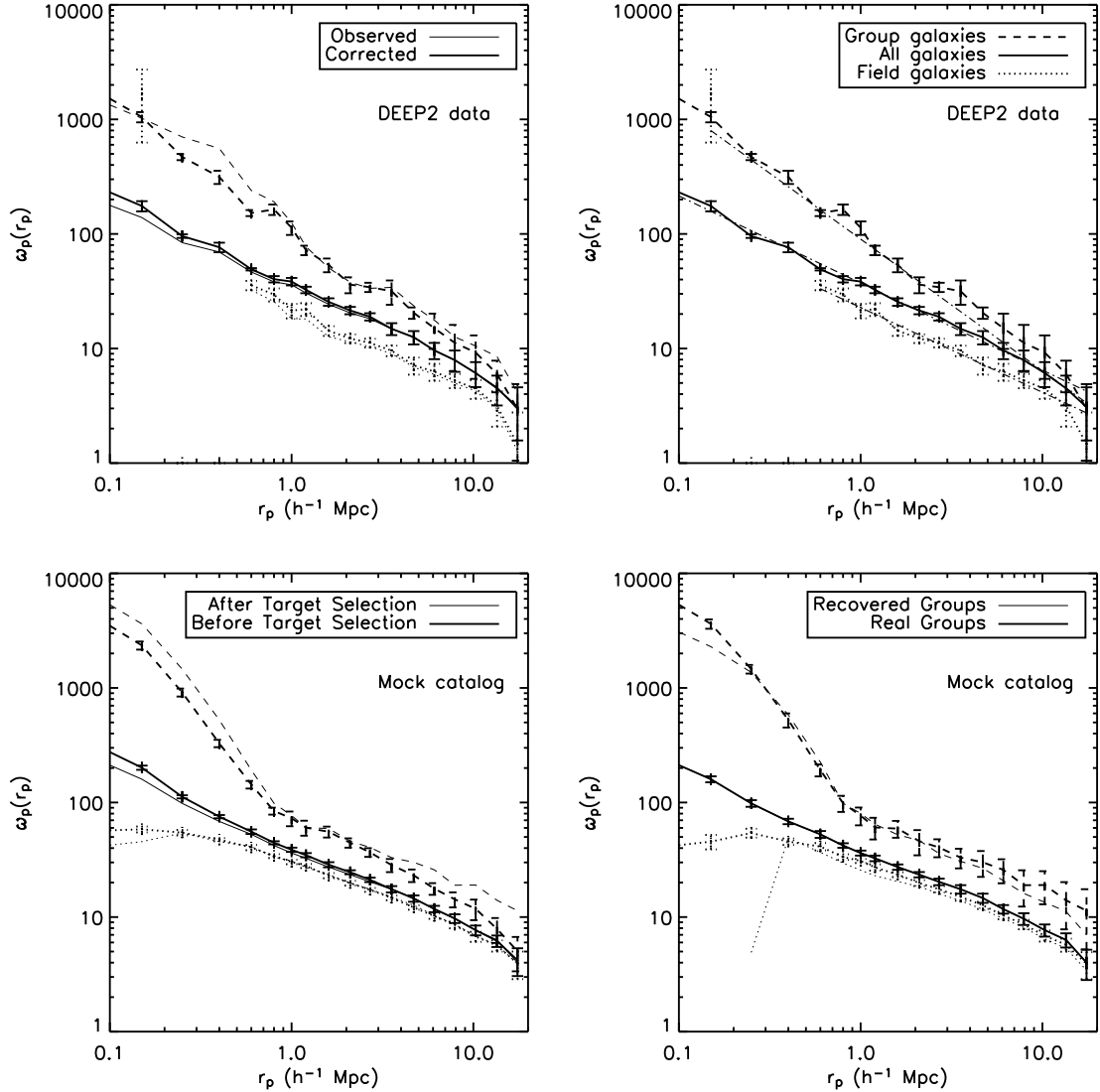}}}}
\caption{
Projected two-point correlation function, \wprp, for the full galaxy 
sample (solid lines)  
and for group galaxy (dashed lines) and field galaxy (dotted lines) 
samples in both the DEEP2 data (top) and
mock catalogs (bottom).  Top left:  Observed correlation functions are shown 
as thin lines without error bars; thick lines with error bars include
corrections for the slitmask target selection algorithm and, for
galaxies in groups or in the field, corrections for the group-finder.  
Top right: Power-law fits to the corrected \wprp \ 
are shown as dot-dashed lines and are given in Table 2. 
Bottom left: The effects of our slitmask target selection are shown
for galaxies in the mock catalogs, before (thin lines without error bars) 
and after (thick lines with error bars) applying the target selection 
algorithm, where we have used real groups to identify group and field galaxies.
  Bottom right:  The effects of the group-finder in the mock catalogs are 
shown for galaxies identified as belonging to real or recovered groups and 
for field galaxies (there is no difference for the full galaxy sample).
\label{mockfield}}
\end{figure}

Fig. \ref{mockfield} presents \wprp \ for 
the full galaxy sample (solid lines), galaxies in groups 
(dashed lines) and for field galaxies (dotted lines), in both 
the DEEP2 data (top) and in the mock catalogs (bottom).  The top left panel
 compares \wprp \ as measured in the observed DEEP2 data (thin
lines with no error bars) with \wprp \ after correcting for effects
due to target 
selection and the group-finder (thick lines with error bars).  To make
this correction, we have used the ratio of \wprp \ as a function of
scale in the mock catalogs for group galaxies, field galaxies, and all
galaxies separately, identified using real groups before target
selection (thick lines with error bars in the bottom right panel), 
to \wprp \ for galaxies identified using 
recovered groups after target selection (thick lines with error bars
in the bottom left panel).  
Power-law fits to the
corrected \wprp \ points are shown in the top right panel of 
Fig. \ref{mockfield} and are listed in table \ref{r0table}, along 
with fits to the observed points.  
Field galaxies are well fit by a power-law on scales $r_p=1-20$
\mpch.  On smaller scales the correlation function is negative,
as field galaxies are not found within $\sim0.6$ \mpch \ of each
other.  Pairs of galaxies within that distance are likely to be part
of a group.  

We will present updated results for \wprp \ for galaxies in the DEEP2 dataset 
as a function of galaxy color, luminosity, redshift, etc. in a future paper 
(Coil et al. 2005, in preparation). Here we focus on the difference 
between the 
clustering of galaxies in groups relative to the full galaxy sample 
(throughout this paper we use the terms ``all galaxies'' and 
``the full galaxy sample'' interchangeably; they include both field and 
group galaxies). 
The full galaxy sample used here includes a total of 9787 galaxies 
in the redshift range $0.7\leq z \leq1.0$ in the same six pointings 
used for the group analysis.  This represents a great advance over 
the sample used in \cite{Coil03xisp}, which contained 2219 galaxies 
in one pointing
over the redshift range $0.7\leq z\leq1.35$.  The fits for $r_0$ and $\gamma$ 
here agree with our earlier results, but the errors, both Poisson and cosmic 
variance, are much smaller in the current sample.  For example, 
we can now address whether the significant rise in slope of 
\wprp \ on small scales predicted by \citet{Kravtsov04}
for galaxies at $z=1$ based upon their simulations is present in the
DEEP2 data.  They find that for a galaxy sample with a comoving number
density of $n=1.5 \times 10^{-2} \ h^3 \ {\rm Mpc}^{-3}$  (similar to
the DEEP2 sample at $z\sim0.8$; see \cite{LLin04})
that the slope of \xir \ changes from
$\gamma\sim1.65$ when measured on scales of $r=0.3-10$ \mpch, where 
$r_0$ is measured to be $\sim4$ \mpch, to
$\gamma\sim1.9$ over scales $r\sim0.1-10$ \mpch, where $r_0\sim3.5$ \mpch.
Fitting our corrected \wprp \ for the full galaxy sample over the
same range in $r_p$ results in $r_p=0.3-10$ \mpch, $r_0=3.64
\pm0.07$ \mpch \ and $\gamma=1.73 \pm0.04$, while for scales $r_p=0.01-10$
\mpch, $r_0=3.64 \pm0.07$ \mpch \ and $\gamma=1.73 \pm0.03$, 
with no change at all in either amplitude or slope.  Therefore no evidence is found 
for a rise in the slope on small scales as predicted by \cite{Kravtsov04}. 
We also do not find a significant difference in the 
slope on scales $r_p<1$ \mpch \ compared to scales $r_p>1$ \mpch, as
is seen for the galaxies in groups.
Fitting for a power-law on scales $r_p=0.05-1$ \mpch, 
$r_0=3.52 \pm0.16$ \mpch \ and  $\gamma=1.78 \pm0.06$, while on
 scales $r_p=1-20$ \mpch, $r_0=3.70 \pm0.10$ and $\gamma=1.77 \pm0.06$.
The full galaxy sample therefore appears to be consistent with a
single power-law slope over the range $r_p=0.1-20$ \mpch.
We note that in the mock catalogs, the slope measured for \wprp \ is 
consistent with the data, though the amplitude of $r_0$ is $\sim7$\% higher.
As stated before, the mock catalogs were designed to match the previously 
published measurements of \wprp \ for the full DEEP2 galaxy sample.

The correlation function for galaxies in groups is relatively well-fit
by a power-law over all scales; a broken power-law fit with a break at
$r_p=1$ \mpch \  results in a steeper slope on small 
scales, with low significance.  The slope for a single
power-law is $\gamma=2.12 \pm0.06$ for scales $r_p=0.05-20$ \mpch,
while it increases to $\gamma=2.16 \pm0.11$ for scales $r_p=0.05-1$
\mpch \ and decreases to $\gamma=2.02 \pm0.15$ for scales $r_p=1-20$
\mpch.  Note the significantly different shape of \wprp \ for group
galaxies in the mock catalogs, which exhibit a strong break at
$r_p\sim1$ \mpch.  This sharp rise on small scales is not seen in the
DEEP2 data.  As we will show in the next subsection, this difference
between the mock catalogs and the data is not due to any systematic
effect from our slitmask algorithm or in our group identification, as the
general shape of the correlation function in the mock catalogs is not
changed by these.  
This difference in the clustering of galaxies in groups between the mock
catalogs and the data is the first indication that the mock catalogs,
which were constructed to match the $z\sim1$ 
luminosity function and clustering of 
all galaxies in the DEEP2 data, do not reproduce additional properties
of the data.  
It appears that the spatial distribution of galaxies in groups is  
less concentrated in the data than in the mock catalogs.
We discuss the implications of this in \S 9.

\begin{figure}
\centerline{\scalebox{0.7}{\rotatebox{90}{\includegraphics{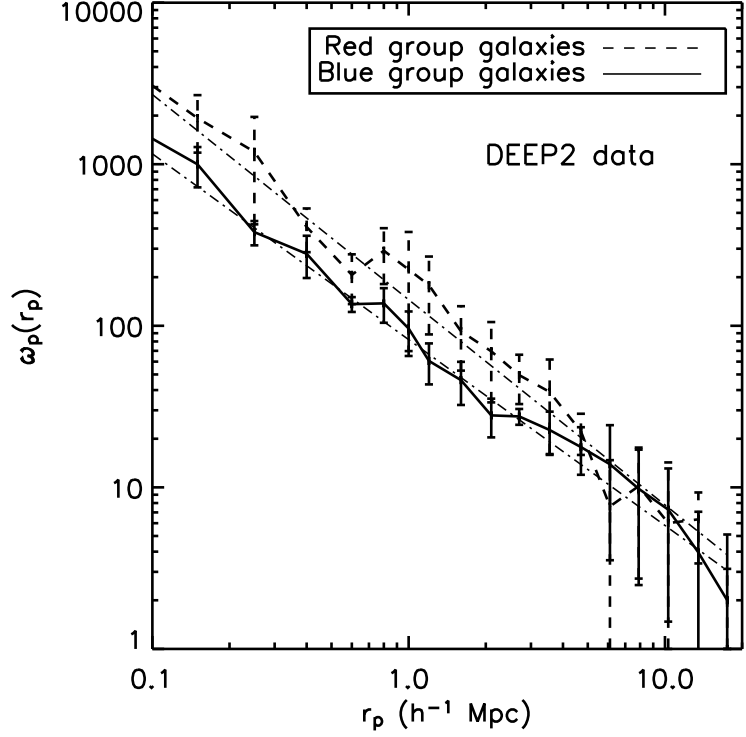}}}}
\caption{Projected two-point correlation function, \wprp, for DEEP2 
galaxies in groups, where we have split the galaxy sample by color using the 
observed bi-modality in restframe $(U-B)_0$.  
Corrections have been applied for effects due to our slitmask
target algorithm and group-finder, using the ratio of \wprp \ in the
mock catalogs for galaxies identified as belonging in real groups
before target selection (thick dashed line in bottom left panel of
Fig. 6) compared to galaxies in recovered groups after
target selection (thin dashed line in bottom right panel in Fig. 6). 
Red group galaxies (dashed line) 
as more strongly clustered than blue group galaxies (solid line) and show a
steeper clustering slope. 
Power-law fits are plotted as thin dash-dot lines and given in Table 2.
\label{grpgalcolor}}
\end{figure}

We also investigate \wprp \ for group galaxies as a function of color.
Fig. \ref{grpgalcolor} shows \wprp \ for galaxies in the DEEP2 data
which are identified as belonging to groups and have red or blue
colors, defined by the minimum in the color bi-modality in restframe
$(U-B)_0$, at $(U-B)_0=1.05$.  In the group galaxy sample, 20\% of
group galaxies are red, while for the full galaxy sample 15\% of
galaxies are red.  Corrections for target selection and our 
group-finder are made using the mock
catalogs as above. Red galaxies in groups have a
steeper slope in \wprp \ and a higher correlation length
than blue galaxies; power-law fits result in $r_0=4.77 \pm0.20$ \mpch
\ and $\gamma=2.15 \pm0.05$ for blue group galaxies and $r_0=5.81
\pm0.45$ \mpch \ and $\gamma=2.27 \pm0.11$ for red group galaxies.
The steeper slope for the red galaxy sample implies that red galaxies are 
more centrally concentrated in their parent dark matter halos; we 
investigate this more directly in \S6.
Colors are not currently included in the mock catalogs
and so we can not present this measurement for the mock catalogs.

\subsection {Effects of Slitmask Target Selection and Group-finder}

As for the clustering of groups in the DEEP2 data, mock catalogs are
used to quantify systematic errors in our measurements of \wprp \ for
each galaxy sample, where real groups are used to define galaxies in
groups or in the field.  
The bottom left panel of Fig. \ref{mockfield}
shows \wprp \ measured for all galaxies, group galaxies, and field
galaxies in the mock catalogs, before (thin lines without error bars)
and after (thick lines with error bars) target selection.  Separate
power-law least-squares fits to \wprp \ in the mock catalogs are
performed for the full galaxy sample and for galaxies in groups and in
the field; the results before and after target selection are shown in
Table \ref{r0table}.  Field galaxies are only affected by the target
selection on the very smallest scales, $r_p\lesssim 0.2$ \mpch, and
there is no significant change to the correlation function of
field galaxies on the scales over which we measure a power-law,
$r_p=1-20$ \mpch.  For the full galaxy sample, our target selection
algorithm causes \wprp \ to be slightly underestimated on small scales
($r_p\lesssim 2$ \mpch) due to our inability to target all close
neighbors.  Before target selection is applied, a power-law fits well
for the full galaxy sample on scales $r_p=1-20$ \mpch, with a steeper
slope on small scales, $r_p=0.1-1$ \mpch.  This change in slope on
small scales in the mock catalogs is more significant (2.5$\sigma$ vs
1.3$\sigma$) for the sample before target selection than after, due to
the smaller error bars for the larger sample.  We note that the scale 
at which the slope changes is larger than predicted by \cite{Kravtsov04} in 
their simulations, and that this difference in slope is not seen in the DEEP2 
data, as discussed above.

For galaxies in groups, however, the effects of target selection are 
more complicated; it {\it increases} their observed clustering on all
scales. In further tests, we find that target selection does not affect the
measured clustering of galaxies in mock catalogs known to belong in
real groups, so long as the group membership is identified before target 
selection.
However, galaxies can be identified as belonging to a group after target
selection only if two or more observed galaxies are in that group.  We
can identify only half of the groups after target selection that were 
detectable 
before target selection, and the groups which are preferentially lost
are those with a low richness.  This causes \wprp \ for
galaxies in groups to be greater when groups are identified after target
selection, as only galaxies in the richest, and presumably highest mass and 
most clustered, groups will be included in the observed sample.

The mock catalogs are also used to investigate the effect of our
group-finder on the measured clustering of galaxies in groups, as we
want to be sure that our group-finding algorithm is not imposing (even
indirectly) a preferred scale for groups, which could potentially
artificially cause a change in slope at small scales in \wprp \
for group galaxies.  This is tested by comparing the clustering of
galaxies in real and recovered groups in the mock catalogs, 
where both samples have had the DEEP2 target selection algorithm applied.
The results are shown in the bottom right panel of Fig.
\ref{mockfield}, where both real and recovered groups are seen to have
an inflection in \wprp \ such that the slope rises on small scales
($r_p<1$ \mpch).  the group-finder, then, is not imposing this scale
on the clustering results.  We also find that the group-finder 
effectively cancels much of the effect of our slitmask target selection, in 
that \wprp \ for galaxies in recovered groups after target selection 
is similar to \wprp \ for galaxies in real groups before target selection.
The overall correction applied to the observed \wprp \ for group galaxies 
in the DEEP2 data is therefore small.

Throughout the paper we compare \wprp \ for the DEEP2 data with mock
catalogs by applying corrections to the observed correlation function
to account for effects of our slitmask target selection and
group-finder and compare to results in the mock catalogs before target
selection for real groups.  None of our results change if instead we 
compare results for the observed \wprp \ in the data to results in the 
mock catalogs before target selection for real groups.

\subsection{One- and Two-Halo Terms of the Group Galaxy Correlation Function}

\begin{figure}
\centerline{\scalebox{0.6}{\rotatebox{90}{\includegraphics{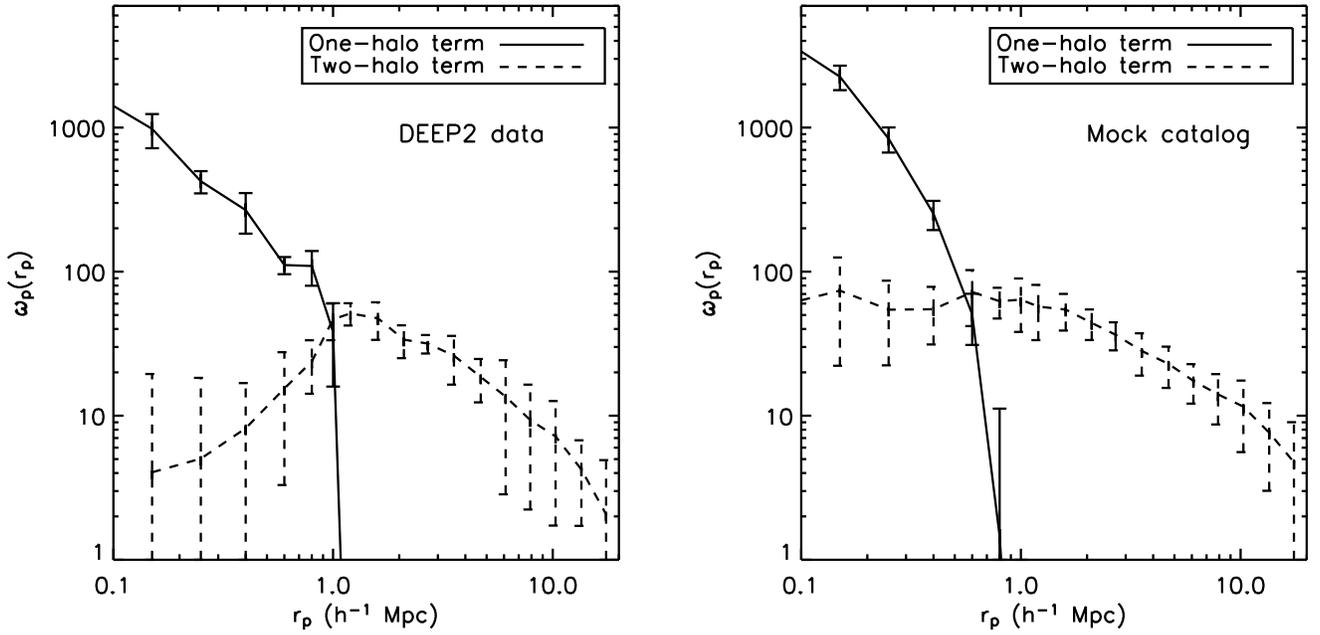}}}}
\caption{
Projected two-point correlation function, \wprp, for galaxies 
in groups in the DEEP2 data (left) and real groups in the 
mock catalogs (right), where the solid and dashed lines include  
pairs inside the same group (`one-halo') or between groups (`two-halo').  
Left: Corrections have been applied for our slitmask target
selection algorithm and group-finder.  The scale at which the two
terms intersect is somewhat larger in the data than in the mock catalogs, 
where the amplitude of the `one-halo' term is significantly higher.
\label{mockseparate}}
\end{figure}

As it is known which group each of the group galaxies belongs to, we can
empirically measure the contribution to \wprp \ 
from pairs of galaxies in the same or in different groups.  This is
akin to determining the `one-halo' and `two-halo' terms of \wprp \ in
the halo model language, where each group is identified with a
single dark matter halo. The total correlation function is then the
sum of these two terms:
\begin{equation}
\xi(r)=[1+\xi_{1h}(r)]+\xi_{2h}(r).
\end{equation}
We calculate the `one-halo' and `two-halo' correlation functions using
the following estimators:
\begin{equation}
\xi_1=\frac{1}{RR}\left[DD_1 \left(\frac{n_R}{n_D}
\right)^2-2DR\left(\frac{n_R}{n_D} \right)+RR\right]
\end{equation}
\begin{equation}
\xi_2=\frac{1}{RR}\left[DD_2 \left(\frac{n_R}{n_D}
\right)^2-2DR\left(\frac{n_R}{n_D} \right)+RR\right],
\end{equation}
where $DD_1$ and $DD_2$ are pair counts of galaxies within the same 
group and in different groups, respectively.  
We then sum these along the line of sight (in the $\pi$ direction, to $\pi_{max}$)
to obtain $w_{p,1h}$ and $w_{p,2h}$.
The projected correlation functions sum such that
\begin{equation}
w_p(r_p)=[\pi_{max}+w_{p,1h}(r_p)]+w_{p,2h}(r_p).
\end{equation}
Fig. \ref{mockseparate} shows $w_{p,1h}(r_p)$ and $w_{p,2h}(r_p)$ for
galaxies in groups in both the DEEP2 data (left) and for real groups 
in the mock catalogs (right).
The data have been corrected for effects due to our target slitmask and
group-finder algorithms.   

In the DEEP2 data the scale at which the `one-halo' and `two-halo'
terms intersect is
$r_p=1.0$ \mpch; the scale in the mock catalogs is $r_p=0.5$ \mpch.  
Exactly where the break occurs between the `one-halo'
and `two-halo' terms will presumably depend on the type of groups we
are probing; larger groups may have this break at a larger radius.
The change in slope seen in \wprp \ for group galaxies in the mock
catalogs is easily understood as the scale at which the `one-halo' and
`two-halo' terms equally contribute to \wprp.  However, the `one-halo'
term has a very different shape in the DEEP2 data than in the mock
catalog, such that galaxies in the data which belong to the same group 
are not as clustered on small scales as in the mock catalogs.  This is
likely due the mock catalogs having an incorrect spatial distribution of 
galaxies within dark matter halos on small scales; we discuss this 
further in \S 9.

\cite{Yang05a} measure \wprp \ for group galaxies in 2dF Galaxy Redshift 
Survey data at $z\sim0$ and find that it is not well fit by a single power-law; 
the `one-halo' term is enhanced relative to the `two-halo' term and 
there is a rise in \wprp \ on scales $r_p\sim1-2$ \mpch.  The strength of
the rise depends on the abundance and luminosities of
the groups; galaxies in more luminous (and presumably more massive) 
groups have a larger `one-halo' term and a stronger
rise in the slope of \wprp \ on small scales.  It is only for the 
full galaxy population (including field galaxies) that they find a single
power-law fit to \wprp.  For our group sample at $z\sim1$, the shape for  
\wprp \ for galaxies in groups is similar to what is seen by 
\cite{Yang05a} for groups with a comparable number density. 
\cite{Yang05a} find a small rise in the slope of \wprp \ on scales 
below $r_p=1$ \mpch \ but do not quantify this.  
Our results at $z\sim1$ appear to be similar to their 
findings at $z\sim0$, though with larger errors due to our smaller sample size.

\section{Group-Galaxy Cross-Correlation Function}

In this section we present the cross-correlation function between 
group centers and the full galaxy sample, which is sensitive to the radial
profile of galaxies in and around groups.  
As with the group and galaxy correlation
functions, to avoid redshift-space distortions we measure the
projected cross-correlation, \wprp. As discussed in \S 2, errors
in the positions of group centers will have some effect on scales
$r_p<1$ \mpch; for this reason we do not plot results for $r_p<0.3$
\mpch.  However, comparisons between data and mock catalogs are
unaffected, as the mock catalogs have been treated in an identical
manner as the data.  

\begin{figure}
\centerline{\scalebox{0.7}{\rotatebox{90}{\includegraphics{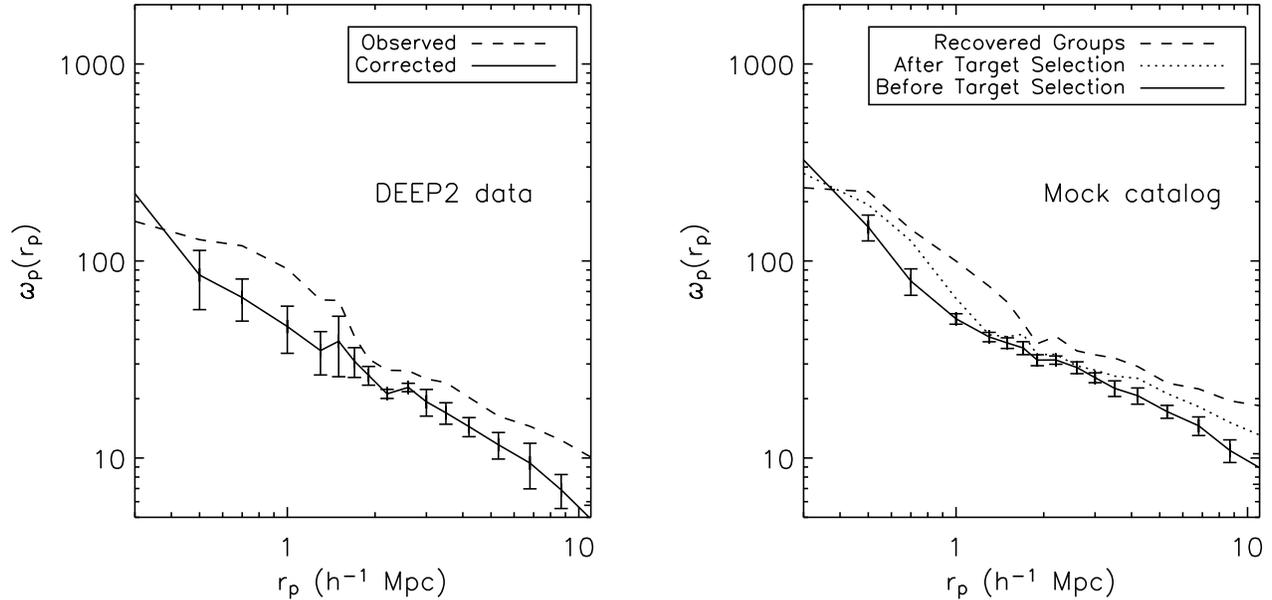}}}}
\caption{Projected cross-correlation function, \wprp, between 
group centers and all 
galaxies in the DEEP2 data (left) and in mock catalogs (right).
In the left panel, the dashed line shows the observed cross-correlation 
function, while the solid line shows the function after correcting for 
effects due to target slitmask and the group-finder.  
In the right panel, the solid and dashed lines
show the cross-correlation function for real groups before 
slitmask target selection is applied and recovered groups after the target
selection is applied, 
 while the dotted line shows the cross-correlation for 
real groups after target selection.  The shape of the cross-correlation in the 
data agrees well with the mock catalogs, though the amplitude is lower.
\label{crossdata}}
\end{figure}

Fig. \ref{crossdata} shows the projected cross-correlation between 
group centers and the full galaxy sample in the DEEP2 data (left) and the mock
catalogs (right).  The left panel shows the observed \wprp \ as a dashed line
and the corrected \wprp \ as a solid line, where corrections for 
target selection and the group-finding algorithm as a function of
scale are made using the ratio of \wprp \ in the mock catalogs 
between real group centers and all galaxies 
 before (solid line, right panel) 
target selection and between recovered group centers
 and all galaxies after (dashed line, right panel) target selection. 
The dotted line in the right panel shows 
the cross-correlation between real groups and all galaxies after
target selection.  
The target selection algorithm has the effect of increasing
the cross-correlation on scales $r_p>0.4$ \mpch, while
decreasing the amplitude on smaller scales. The small-scale
decrease is due to our inability to target galaxies which are in close
projection on the plane of the sky; this causes us to undersample
close neighbors.  On large scales, the effect is due to the slitmask
target algorithm affecting {\it which} groups we identify; after
target selection we lose many of the pairs of galaxies which were
identified as groups before such that we preferentially identify the groups
with more observed members, which are presumably more massive and 
therefore more clustered. The effect of the group-finding algorithm is
to increase the cross-correlation on scales $r_p>0.4$ \mpch,
where the group-finder has by definition targetted overdensities in
the galaxy distribution.

Comparing the solid lines in the two panels of Fig. \ref{crossdata}, which 
shows \wprp \ for real groups before target selection to the corrected data,
the overall shape of the cross-correlation agrees reasonably well, 
 though the amplitude is somewhat higher in the mock catalogs 
on both small and large scales.  This is consistent with
what is seen for the correlation function of galaxies in groups shown in
Fig. \ref{mockfield}, which are more strongly clustered in the mock catalog 
than in the DEEP2 data.

\begin{figure}
\centerline{\scalebox{0.7}{\rotatebox{90}{\includegraphics{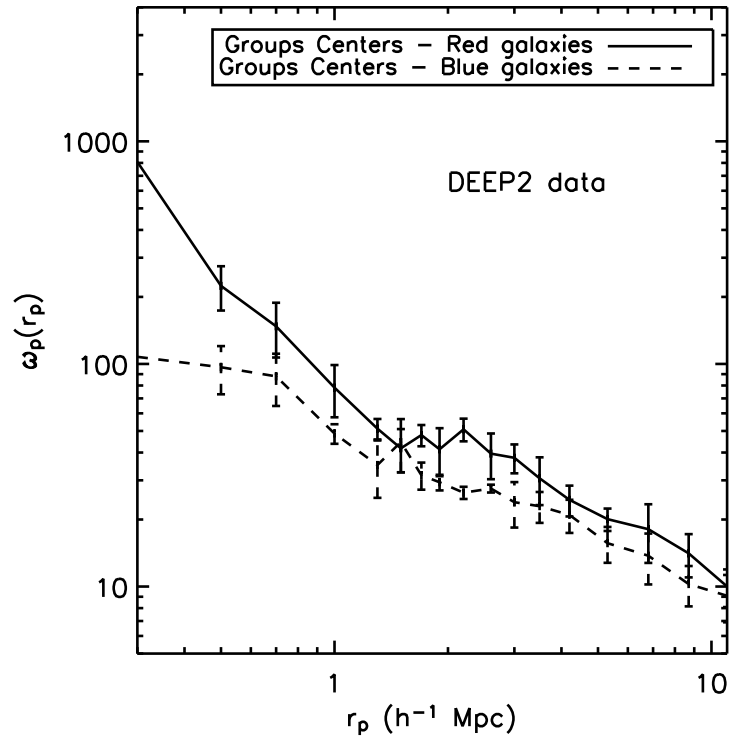}}}}
\caption{Projected cross-correlation function, \wprp, between group centers 
and the full galaxy sample in the DEEP2 data, for galaxies redward (solid 
line) and blueward (dashed line) of the observed bi-modality in restframe 
$(U-B)_0$ color. Corrections have been applied for our slitmask target
selection algorithm and group-finder. 
Red galaxies are found preferentially near the centers 
of groups at $z\sim1$. 
\label{redcross}}
\end{figure}

We also investigate the dependence of the radial distribution of galaxies
in groups on galaxy color. Fig. \ref{redcross} shows the projected
cross-correlation function between either red or blue galaxies and group
centers, where again the galaxy sample has been split at the bi-modality in
the restframe $(U-B)_0$ color distribution at $(U-B)_0=1.05$.  Within groups, 
on small scales,
$r_p\lesssim0.5$ \mpch, red galaxies are much 
more strongly clustered than blue 
galaxies, i.e., red galaxies are preferentially found near the
centers of groups.  In the DEEP2 data 
20\% of group galaxies in our sample are red, while 13\% of field galaxies 
and 15\% of the full galaxy sample (used in this cross-correlation) are red.

Similar trends are seen at $z\sim0$ by \cite{Yang05b}, who measure the
cross-correlation between group centers and all galaxies in 2dF and
SDSS data.  They also find a difference between the radial distribution of 
red and blue galaxies, though it is only apparent for groups with
masses $M\gtrsim 10^{13} h^{-1} M_\sun$, and the differences are 
smaller than those found here at $z\sim1$.

\section{Relative Bias Between Groups and Galaxies}

\begin{figure}
\centerline{\scalebox{0.9}{\rotatebox{90}{\includegraphics{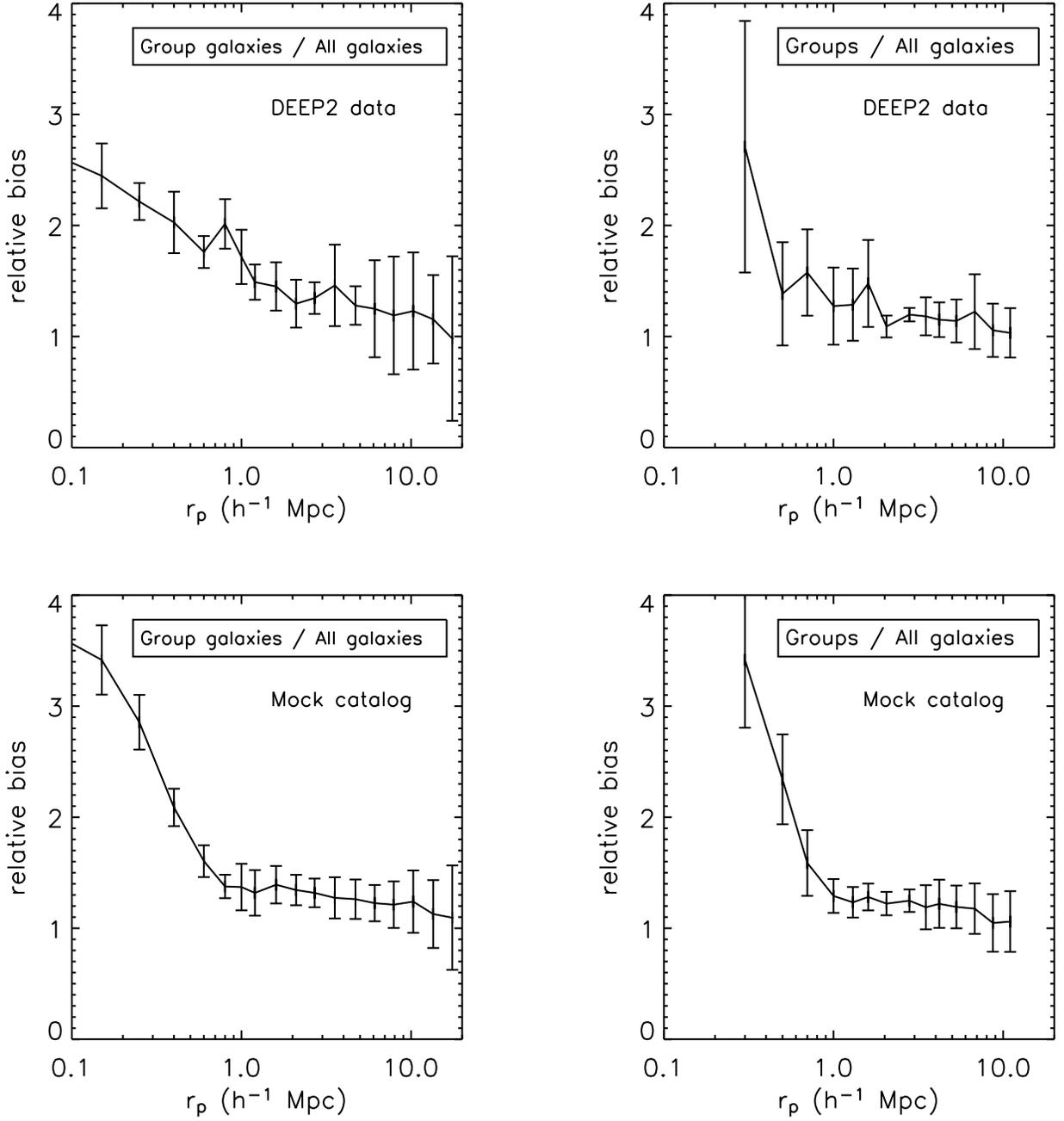}}}}
\caption{Scale-dependence of the relative bias between galaxies in groups 
and the full galaxy sample (left) and between groups and galaxies
(right) in the DEEP2 data (top) and mock catalogs (bottom).
The data do not show nearly as strong of 
a scale-dependence in the relative bias between group galaxies and all galaxies
on small scales (upper left) as is seen in the mock catalogs (bottom left).  
The relative bias between
groups and galaxies in the data (upper right) agrees reasonably 
well the mock catalogs (bottom right).
\label{grpgalbias}}
\end{figure}

Measuring the clustering properties of groups, all galaxies, and
galaxies in groups in the DEEP2 data allows us to measure the
relative bias between galaxies in groups and all galaxies and 
 between groups and galaxies.  Fig. \ref{grpgalbias} plots the
relative bias of group galaxies to the full galaxy sample, 
which we define as the square
root of \wprp \ for group galaxies (dashed lines in Fig. 6) divided
by \wprp \ for the full galaxy sample (solid lines in Fig. 6), as a function
of scale for $r_p=0.1-20$ \mpch \ in both the DEEP2 data (top left
panel) and the the mock catalogs (bottom left panel), after correcting
for target selection and the group-finder. 
The bias seen between group galaxies and all galaxies is not surprising, as 
group galaxies reside in more massive dark matter halos.  There is a
clear scale-dependence to the relative bias between group galaxies and
all galaxies in the DEEP2 data, which falls from $b_{rel}\sim2.5 \pm0.3$
at $r_p=0.1$ \mpch \ to $b_{rel}\sim1 \pm0.5$ at $r_p=10$ \mpch.  The
mock catalogs have a much higher relative bias on small scales
($r_p\lesssim1$ \mpch) which does not match the bias seen in the data.
This reflects the strong rise in slope of the correlation function of
galaxies in groups seen on small scales in the mock catalogs.

The ratio of the group center-full galaxy sample 
cross-correlation function (Fig. 9) to the galaxy correlation
function (solid lines in Fig. 6) provides a measure of the relative
bias of groups to galaxies, which is shown on the right side of Fig.
\ref{grpgalbias}.  We have corrected for slitmask target selection and
the group-finder. There is some scale-dependence to the relative bias
between groups and galaxies in the DEEP2 data (top right panel) and 
the weighted mean relative bias is $b_{rel}=1.17 
\pm0.04$ over scales $r_p=0.5-15$ \mpch.  
The mock catalogs have a mean value of $b_{rel}=1.23 \pm0.02$
for $r_p=0.5-15$ \mpch, in reasonable agreement with the data, though the
mock catalogs again show a higher bias on small scales, below
$r_p\sim1$ \mpch, and the agreement with the data is better on scales
$r_p>1$ \mpch.  These measures of the
relative biases of groups to galaxies and galaxies in groups to all
galaxies at $z\sim1$ are further constraints which simulations and
galaxy evolution models must match, in addition to measures of \wprp \
for all galaxies and for groups.  We discuss the implications of these
differences between the data and mock catalogs in the next two sections.

\section{Effect of Varying the Halo Model Parameters}

\begin{figure}
\centerline{\scalebox{0.7}{\rotatebox{90}{\includegraphics{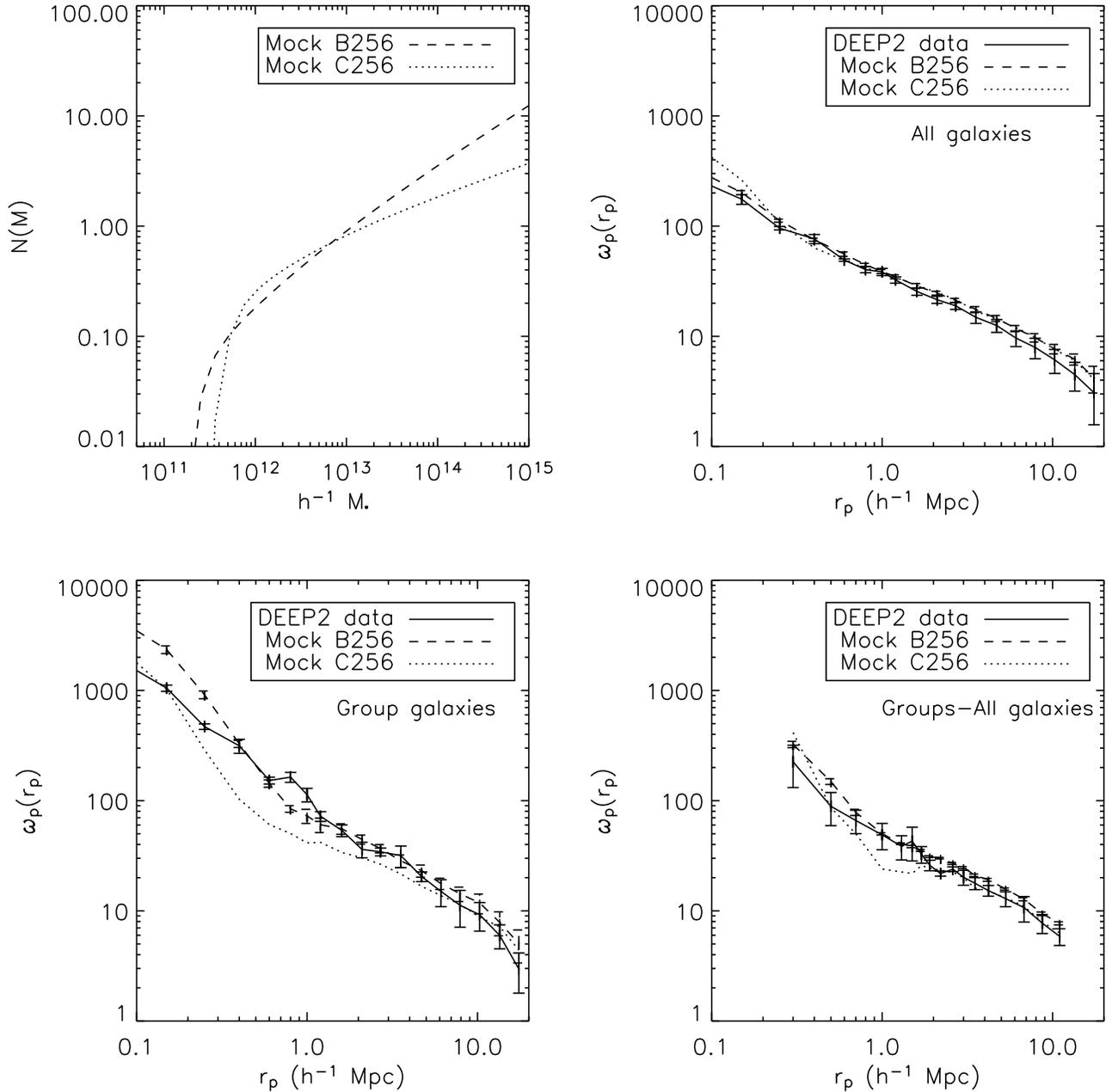}}}}
\caption{The effects of varying the halo occupation distribution (HOD)
parameters on various clustering measures.  
Top left: HODs for the two mock catalogs analyzed here.  Shown is the 
average number of galaxies with $L>L^*$ placed in 
dark matter halos, as a function of the halo mass. 
The projected correlation function for all galaxies (top right) is similar
in both of the mock catalogs (dotted and dashed lines), which agree
 fairly well with the DEEP2 data (solid line).
The projected correlation function for galaxies in groups (bottom
left) shows a rise on small scales in  
both mock catalogs that is not seen in the DEEP2
data.  Varying the HOD does not appear to resolve this discrepancy.  
The projected cross-correlation function between group centers 
and all galaxies (bottom right) agrees fairly well between the mock catalogs 
and data.  
\label{grpgalhod}}
\end{figure}

The differences seen between the clustering of group galaxies on small
scales in the DEEP2 data and the mock catalogs could be due to the
mock catalogs having the wrong spatial distribution for galaxies within
their parent dark matter halos and/or the wrong HOD, which specifies 
the probability that a
dark matter halo of mass M hosts N galaxies, $P(N|M)$.  To illustrate how
much these differences may be due to the HOD used to create the mock
catalogs, we investigate the clustering of group galaxies in two mock
catalogs with similar number densities and different HODs. In addition
to the mock catalog used throughout this paper (labeled as 
``B256''), we also analyze a mock catalog in which a different HOD was
applied to the same dark matter simulation; this model is labeled as
``C256'' and was chosen as one of the most discrepant HOD models that
has an observed \wprp \ for the full galaxy sample that, by design, 
matches the results for the DEEP2 data published in \cite{Coil03xisp}.
The HODs for galaxies with $L>L*$ for these two models are shown in 
the upper left of Fig.
\ref{grpgalhod}, where model B256 is seen to have a lower minimum halo
mass hosting a single galaxy, and a steeper slope on larger mass
scales ($\sim0.5$ compared to $\sim0.26$ for the C256 model), which
results in having a greater fraction of galaxies residing 
in massive halos. The curves for galaxies with lower luminosity thresholds 
have a similar shape and higher amplitude than what is shown here 
(see Fig. 1 in \cite{Yan03}).  Mock catalogs made
with the C256 model have 35\% of galaxies in recovered groups after
target selection, similar to what is found for the DEEP2 data (39\%),
and higher than the value found in the B256 model (27\%), even though
the C256 model has relatively fewer galaxies in more massive halos. 
This is due to the parameters of the group-finding algorithm having 
been tuned to match the observed $n(\sigma,z)$ of the DEEP2 data for 
$\sigma\ge350$ \kms; we have not re-tuned the group-finder to the C256
mock catalogs or our $\sigma\ge200$ \kms \ cutoff.  Both the B256 and C256 mock
catalogs have the same number density for the full galaxy sample.   
The clustering measures shown here have all been corrected for slitmask
target effects and the group-finder. 

Fig. \ref{grpgalhod} shows the correlation function for
all galaxies (top right panel) and for group galaxies (bottom left
panel) in each of the two halo model mock catalogs.  The \wprp \ for
the full galaxy sample is very similar in the two catalogs; the only
differences are on scales less than $r_p\sim0.5$ \mpch, where the C256
model has a slightly higher clustering amplitude.  For \wprp \ for
group galaxies, the overall shape is similar for the two models but
the amplitude in model B256 is higher at all scales, as this model has
more galaxies in massive halos, such that the group galaxies will be
more clustered.  These figures show that the amplitude of the group
galaxy correlation function adds an additional constraint on the
HOD, which is not gained from the correlation function of all
galaxies alone.  It also shows that the general shape of the group
galaxy correlation function, and in particular, the rise in slope
on small scales, is {\it not} sensitive to the parameters of the HOD
used.

The bottom right panel of Fig. \ref{grpgalhod} presents the
cross-correlation function of group centers and all galaxies in both
mock catalogs.  Here there is a difference in the shape of the
cross-correlation on small scales for the different HODs. The C256
model has a lower amplitude than the B256 model over almost all scales
but shows a distinct rise on the smallest scales, $r_p\lesssim0.5$
\mpch, which is not seen in the B256 model.
Indeed, both the correlation function for all galaxies and the 
group-galaxy cross-correlation function in the C256 model show a rise 
on small scales 
that is not seen in the B256 model; this results from the C256 model having
preferentially more galaxies in smaller mass halos which dominate the 
pair counts at small separations.

Results from the DEEP2 data are also compared to the different halo model
mock catalogs in Fig. \ref{grpgalhod}.  By design, \wprp \ for all 
galaxies matches both 
mock catalogs well, though the data do not show the rise on the smallest scales
that is seen for the C256 model. The shape of the correlation function for
group galaxies in the DEEP2 data does not match either of the halo
model mock catalogs; the data show a significantly 
shallower slope on small scales.
The amplitude of the cross-correlation function agrees better with the B256
model than the C256 model, and the shape of the cross-correlation 
disagrees with the C256 model on small scales.  
The significant difference in \wprp \ 
for group galaxies on small scales is presumably due
to a difference in the spatial distribution of galaxies in groups in the data
and the mock catalogs, as it does not appear to be reconcilable by
altering the HOD.  This implies that the spatial distribution
of galaxies in groups in the mock catalogs is incorrect.  This will be
discussed further in the next section.

We note that if the C256 mock catalogs had been used to correct the
observed \wprp \ for the full DEEP2 galaxy sample and 
galaxies in groups as presented
in Fig. \ref{mockfield} for slitmask target effects, none of our conclusions 
in the paper would change, as the relative differences before and after target
selection are similar in the two mock catalogs.  The differences
for all DEEP2 galaxies if using the C256 mock catalogs to correct for
target selection effects are negligable, well within the $1 \sigma$ errors quoted in 
Table \ref{r0table}.  The differences for group galaxies are within
the $2 \sigma$ errors, with both the corrected $r_0$ and $\gamma$ being lower
($r_0=4.72 \pm0.23$ \mpch \ and $\gamma=2.05 \pm0.06$), and there is
still no significant difference in the slope of \wprp \ on small and large
scales in the DEEP2 data.

\section{Discussion and Conclusions}

Groups bridge the gap between galaxies and clusters in both mass and 
scale, and are also the likely locations of galaxy mergers.  Measurements of
the clustering of groups can constrain cosmological parameters, and 
measurements of the clustering of galaxies in groups can constrain 
both halo model parameters and the spatial profile of galaxies in 
their parent dark matter halos.  Here we present 
the first results on the clustering of groups and galaxies in groups 
at $z\sim1$.
We measure four types of correlation function statistics 
in the DEEP2 dataset: 1) the group correlation function, 2) the
galaxy correlation for the full galaxy sample, 3) the galaxy 
correlation function for galaxies in groups, and 4) the
group-galaxy cross-correlation function.  The first clustering measure
probes the dark matter halo-halo correlation function on mass scales
of galaxy groups, which
is well-understood from dark matter simulations alone.
The clustering of groups in the 
DEEP2 data at $z\sim1$ matches predictions for a \lcdm cosmology and is used
to estimate the typical dark matter masses of the halos the groups
studied reside in.  

The second clustering measure, the galaxy correlation function for the
full DEEP2 galaxy sample, is an update on results using early
DEEP2 data presented in \cite{Coil03xisp}. Here we present this 
measurement for the full galaxy sample using a much larger dataset,
with over four times as many galaxies covering three fields in the
sky; the statistical error on $r_0$ is now 2\%.  We show that \wprp \ for the
full galaxy sample provides constraints on the
halo occupation distribution (HOD), the number of galaxies that reside in a
dark matter halo of a given mass), though not in an unique
way. There is some leeway in how halos can be populated
which results in a correlation function that matches
our measurements, as shown in the upper right panel of Fig. 12.  

The third clustering measure, the correlation function of
galaxies in groups, is similar to the second but is restricted to
galaxies in more massive halos, as measuring the clustering of
galaxies in groups is sensitive to a higher halo mass range than for
the full galaxy sample.  The contribution from the `one-halo' term is
necessarily higher for galaxies in groups as these galaxies are
identified as belonging in halos with several other galaxies.  This
provides further constraints on the halo model parameters than those 
from the measurement of \xir \ for all galaxies alone. 
 We also find that red galaxies in groups have a
steeper slope and higher clustering amplitude than blue galaxies in groups.

The fourth clustering measure, group-galaxy cross-correlation function,  
reflects the spatial distribution of galaxies within dark matter halos
above a given mass, and depends as well upon the parameters of the
halo model.  We find using the group center-galaxy cross-correlation function
 that red galaxies are found preferentially in the centers of
groups compared to blue galaxies, which has also been 
seen locally \citep{Collister05, Yang05b}.  We
find that this trend is in place at $z\sim1$.

All four of these measurements are compared to
mock catalogs constructed from N-body simulations and 
depend differently on, and can therefore
simultaneously constrain, both parameters of the halo model and the
spatial profile of galaxies within halos.
Comparing these clustering measurements in the DEEP2 data with the
mock catalogs of \cite{Yan03}, three of the four measures
agree fairly well with the simulations, with the exception of the 
correlation function for galaxies in groups.  
The clustering amplitude for the full galaxy sample roughly 
matches the mock catalogs; this is by design: the catalogs were constructed 
with an HOD that is
consistent with earlier DEEP2 clustering results for all galaxies.
The HOD used is not uniquely determined however; the observed \wprp \
for the full DEEP2 galaxy sample can be matched with substantially
different HODs (two samples are shown in the upper left panel of
Fig. 12).  We leave an improved HOD reconstruction from \wprp \
for the full galaxy sample for a future paper, where we will study the
clustering as a function of galaxy properties such as luminosity, color,
redshift, etc., in volume-limited samples; here we focus on comparing
the clustering of galaxies in groups to all galaxies and the different
constraints they provide on the HOD.  We do note that \wprp \ for all
galaxies is fit by a single power-law on scales $r_p=0.05-20$ \mpch, 
with $r_0=3.63 \pm0.07$ and $\gamma=1.74 \pm0.03$.

While the mock catalogs have similar
projected clustering for the full galaxy population to the DEEP2 data
by design,
there is a strong discrepancy in the clustering of galaxies in groups.
The DEEP2 data do not show a significant rise in the
slope of \wprp \ on small scales, for either group galaxies or the
full galaxy sample (upper panels of Fig. 6).  
In contrast, our mock catalogs have a very strong rise on
small scales for \wprp \ for group galaxies, though not for the full galaxy 
sample (bottom panels of Fig. 6).  
To test whether this discrepancy can be accounted for by the halo
model parameters used, we analyze mock catalogs constructed with
a different HOD but similar clustering for the full galaxy sample
(Fig. 12).
We find that there is still a rise in slope for the clustering of galaxies
in groups in the second mock catalogs (model C256) which is not seen in the data.
This result is unaffected by our definition of the group center.  

The slope of \wprp \ for group galaxies on small scales should depend quite 
sensitively on the spatial distribution of galaxies within dark matter halos.  
We therefore conclude that there is a difference in the 
spatial distribution of galaxies within their parent dark matter halos
in the DEEP2 data and our mock catalogs.  The mock catalogs assume no
spatial bias, except for the assumption of a central galaxy; 
the most luminous galaxy in a halo is
placed at the center of the halo, while all subsequent galaxies are
assigned to random dark matter particles, following an NFW profile.  
Assuming that the brightest galaxy occupies the very center of the halo 
is likely not to be correct, as groups at $z=1$ are not expected to have 
a large, dominant, bright galaxy in their centers.  This assumption will
result in a higher correlation function on small scales.

In our mock catalogs the satellite galaxies are assumed to 
follow the same NFW profile 
as the dark matter particles; this appears to not be the 
correct spatial profile for the galaxy population at $z=1$.
There is evidence in both simulations and data at $z\sim0$ that
satellite galaxies do not follow the same spatial profile as the dark matter
particles.  Simulations have found that subhalos have a shallower
spatial profile than the dark matter particles at $z\sim0$
\citep[e.g.,][]{Gao04a,Diemand04, Nagai05}, though exactly how
galaxies are related to subhalos is still not entirely known.
Observationally, several authors have measured the spatial profiles of
galaxies in groups in data at $z\sim0$.  Using the Two Micron All Sky
Survey, \cite{Lin04} stack groups and clusters to measure the radial
mass-to-light profile and find that galaxies are less concentrated in
the centers of groups and clusters than the dark matter.
\cite{Collister05} use the 2dF 2PIGG group catalog to directly measure
the radial profile of galaxies within groups and find that galaxies
are less centrally concentrated than what is seen for dark matter
particles in simulations.  Similar results are found by
\cite{Hansen05} for clusters in SDSS, and by \cite{Diaz05} and
\cite{Yang05b} for groups in SDSS and 2dF. 

The cross-clustering between groups and galaxies matches the mock catalogs
well on scales $r_p>0.5$ \mpch; on smaller scales the mock catalogs have
a slightly steeper slope. 
The cross-correlation between groups and galaxies is linearly 
proportional to the radial distribution of galaxies in groups, 
while the correlation of
group galaxies is proportional to the second power of the radial
distribution; this may be why the shape agreement between the data
and the mock catalogs is better for the cross-correlation than for the
correlation of group galaxies.  The group-galaxy 
cross-correlation function can also be affected by uncertainties in the 
location of the center of each group which can dilute the signal on 
small scales, unlike for the correlation function of galaxies in groups.
\cite{Yang05b} find that the cross-correlation between group 
centers and galaxies at $z=0.1$ in 2dF and SDSS data 
is lower on scales $r_p<0.1$ \mpch \ than in their mock catalogs, 
which do not have a
spatial bias with respect to the dark matter distribution.  
They create a series of mock catalogs with NFW profiles for the 
galaxies with lower concentration parameters, $c$, than in the 
dark matter and find that catalogs with concentration values of 
about one-third the value for the dark matter halos match their data well.  
These mock catalogs show the same trend which is required here, namely a lower 
cross-correlation on small scales of $r_p<0.1$ \mpch. 

We show here that the clustering properties of galaxies in groups can
be used to break degeneracies among different HODs that can not be
distinguished by the
clustering of all galaxies alone.  We find that galaxies in the DEEP2
data do not have the same spatial profile as in our mock catalogs,
which assumes a central galaxy surrounded by satellite 
galaxies following an NFW profile.
Using the clustering statistics presented in this paper will allow us
to now construct more realistic mock catalogs for the DEEP2 survey that 
have a better constrained HOD and radial profile for galaxies within 
dark matter halos.

\acknowledgements
We would like to thank the anonymous referee for 
helpful comments and Zheng Zheng for useful discussions.  
This project was supported by the NSF grant AST-0071048.
J.A.N. acknowledges support by NASA through Hubble Fellowship grant
HST-HF-01165.01-A awarded by the Space Telescope Science Institute,
which is operated by AURA Inc. under NASA contract NAS 5-26555. 
C.-P. Ma is supported in part by NASA grant NAG5-12173 and NSF
grant AST-0407351. S.M.F. 
would like to acknowledge the support of a Visiting Miller
Professorship at UC Berkeley.  The DEIMOS spectrograph was funded by a
grant from CARA (Keck Observatory), an NSF Facilities and
Infrastructure grant (AST92-2540), the Center for Particle
Astrophysics and by gifts from Sun Microsystems and the Quantum
Corporation.  The DEEP2 Redshift Survey has been made possible through
the dedicated efforts of the DEIMOS staff at UC Santa Cruz who built
the instrument and the Keck Observatory staff who have supported it on
the telescope.  The data presented herein were obtained at the
W.M. Keck Observatory, which is operated as a scientific partnership
among the California Institute of Technology, the University of
California and the National Aeronautics and Space Administration. The
Observatory was made possible by the generous financial support of the
W.M. Keck Foundation. The DEEP2 team and Keck Observatory acknowledge
the very significant cultural role and reverence that the summit of
Mauna Kea has always had within the indigenous Hawaiian community and
appreciate the opportunity to conduct observations from this mountain.



\clearpage

\begin{deluxetable}{lrcccc}
\tablewidth{0pt}
\tablecolumns{6}
\tablecaption{Power law fits to the group correlation functions 
\xis \ and \xir \ for groups in the
  DEEP2 data and in mock catalogs on scales $r=3-20$ \mpch.}
\tablehead{
\colhead{Sample} & \colhead{No. of} & \colhead{Observed}   &    & \colhead{Corrected} &\\
\colhead{}       & \colhead{groups} & \colhead{$s_0$}    & \colhead{$\gamma$} & \colhead{$r_0$}     & \colhead{$\gamma$} \\
\colhead{}       & \colhead{}       & \colhead{\mpch}    & \colhead{}         & \colhead{\mpch}     & \colhead{}        
}
\startdata
{\bf DEEP2 data} &                  &                    &                    &                     &                    \\
$N\ge2$          & 460              &$7.3 \pm0.6$        &  $1.4 \pm0.2$      &$6.2 \pm0.4$         & $1.5 \pm0.2$       \\ 
$N\ge4$          & 204              &$9.5 \pm0.9$        &  $1.6 \pm0.3$      &$8.3 \pm0.8$         & $1.6 \pm0.3$       \\
{\bf Mock Catalog}&                 &                    &                    &                     &                    \\   
Recovered        & 395              &$7.9 \pm0.5$        & $1.6 \pm0.2$       &                     &                    \\  
Real             & 281              &$7.5 \pm0.8$        & $1.6 \pm0.3$       &$6.0 \pm0.7$         & $1.8 \pm0.3$       \\  
Before Target Selection & 527       &                    &                    &$7.4 \pm0.4$         & $1.6 \pm0.2$       \\  
\enddata
\label{grptable}
\end{deluxetable}

\begin{deluxetable}{lrccccc}
\tablewidth{0pt}
\tablecaption{Power law fits to the galaxy correlation function \wprp \ for galaxies in the DEEP2 data and in mock catalogs.  Observed 
values are after target selection, while corrected values are before target selection.}
\tablehead{
\colhead{Sample} & \colhead{No. of} & \colhead{$r_p$ range} & \multicolumn{2}{c}{Observed} & \multicolumn{2}{c}{Corrected} \\
\colhead{} & \colhead{galaxies} & \colhead{} & \colhead{$r_0$} & \colhead{$\gamma$} & \colhead{$r_0$} & \colhead{$\gamma$} \\
\colhead{} & \colhead{} & \colhead{\mpch} & \colhead{\mpch} & \colhead{} & \colhead{\mpch} & \colhead{}        
}
\startdata
{\bf DEEP2 data} &    &            &                &                &                 &                \\
Group galaxies & 3840 & $0.05-20$  & $5.68 \pm0.26$ & $2.12 \pm0.05$ & $5.26 \pm0.27$  & $2.12 \pm0.06$ \\
               &      & $0.05-1$   & $7.21 \pm0.73$ & $1.96 \pm0.06$ & $5.08 \pm0.62$  & $2.16 \pm0.11$ \\
               &      & $1-20$     & $5.50 \pm0.33$ & $1.95 \pm0.19$ & $5.34 \pm0.30$  & $2.02 \pm0.15$ \\
Red group galaxies & 794 & $0.05-20$ & $6.31 \pm0.44$ & $2.29 \pm0.09$ & $5.81 \pm0.45$ & $2.27 \pm0.11$ \\
Blue group galaxies & 3046 & $0.05-20$ & $5.10 \pm0.20$ & $2.12 \pm0.04$ & $4.77 \pm0.20$ & $2.15 \pm0.05$ \\
All galaxies   & 9787 & $0.05-20$  & $3.46 \pm0.09$ & $1.68 \pm0.04$ & $3.63 \pm0.07$  & $1.74 \pm0.03$ \\
Field galaxies & 5947 & $1-20$   & $2.55 \pm0.26$ & $1.68 \pm0.12$ & $2.76 \pm0.20$  & $1.72 \pm0.12$ \\
&&&&&&\\
{\bf Mock Catalogs} (B256)  & &           &                &                 &                &                 \\
Group galaxies & 3020  & $0.05-1$  & $3.63 \pm0.35$ & $2.84 \pm0.12$  & $3.02 \pm0.25$ & $2.99 \pm0.13$  \\
               &       & $1-20$    & $5.57 \pm0.37$ & $1.75 \pm0.16$  & $5.43 \pm0.29$ & $1.84 \pm0.11$  \\
All galaxies   & 11262 & $0.05-20$ & $3.75 \pm0.09$ & $1.69 \pm0.03$  & $3.75 \pm0.09$ & $1.69 \pm0.03$  \\
               &       & $0.05-1$  & $3.55 \pm0.18$ & $1.76 \pm0.07$  & $3.52 \pm0.14$ & $1.85 \pm0.05$  \\
               &       & $1-20$    & $3.71 \pm0.15$ & $1.65 \pm0.07$  & $3.85 \pm0.11$ & $1.68 \pm0.05$  \\
Field galaxies &  8242 & $1-20$    & $3.01 \pm0.14$ & $1.57 \pm0.06$  & $3.35 \pm0.12$ & $1.63 \pm0.05$  \\

\enddata
\label{r0table}
\end{deluxetable}

\clearpage

\end{document}